\documentclass[final,5p,twocolumn]{elsarticle} 
\usepackage{todonotes}
\journal{Journal of Systems and Software}

\usepackage{natbib}
\usepackage{url}
\usepackage{graphicx}
\usepackage{array}
\usepackage{multirow}
\usepackage{float}
\usepackage{tcolorbox}
\usepackage{fourier} 
\usepackage{caption}
\usepackage{subcaption}
\usepackage{url}

\newcolumntype{L}[1]{>{\raggedright\let\newline\\\arraybackslash\hspace{0pt}}p{#1}}
\newcolumntype{C}[1]{>{\centering\let\newline\\\arraybackslash\hspace{0pt}}p{#1}}
\newcolumntype{R}[1]{>{\raggedleft\let\newline\\\arraybackslash\hspace{0pt}}p{#1}}

\def\BibTeX{{\rm B\kern-.05em{\sc i\kern-.025em b}\kern-.08em
    T\kern-.1667em\lower.7ex\hbox{E}\kern-.125emX}}

\definecolor{gray50}{gray}{.5}
\definecolor{gray40}{gray}{.6}
\definecolor{gray30}{gray}{.7}
\definecolor{gray20}{gray}{.8}
\definecolor{gray15}{gray}{.85}
\definecolor{gray10}{gray}{.9}
\definecolor{gray05}{gray}{.95}

\definecolor{darkmagenta}{rgb}{0.55, 0.0, 0.55}
\definecolor{darkgreen}{RGB}{6, 46, 3}
\definecolor{amber}{rgb}{1.0, 0.75, 0.0}
\definecolor{ao(english)}{rgb}{0.0, 0.5, 0.0}
\definecolor{redmild}{RGB}{234, 67, 53}

\newcommand{\negtd}[1]{\textcolor{redmild}{#1}}
\newcommand{\postd}[1]{\textcolor{ao(english)}{#1}}

\definecolor{arsenic}{rgb}{0.23, 0.27, 0.29}

\definecolor{main}{HTML}{CFCFCF}    
\definecolor{sub}{HTML}{CFCFCF}     
\newcounter{keyTakeAwaysCounter}

\usepackage{fontawesome}
\newtcolorbox{boxC}{
    colback = sub, 
    boxrule = 0pt,  
}

\newenvironment{keyTakeAways}[1][Key Take Away]
    {
    \addtocounter{keyTakeAwaysCounter}{1}
        \begin{boxC}
        \faLightbulbO ~ \thekeyTakeAwaysCounter. \textbf{#1}.\\
        }{
        
        \end{boxC}
}

\begin{document}
\begin{frontmatter}

\title{The Dual-Edged Sword of Technical Debt: Benefits and Issues Analyzed Through Developer Discussions}


\author[OULU]{Xiaozhou Li}
\ead{xiaozhou.li@oulu.fi}
\author[OULU]{Matteo Esposito}
\ead{matteo.esposito@oulu.fi}
\author[UNIBZ]{Andrea Janes}
\ead{andrea.janes@unibz.it}
\author[OULU]{Valentina Lenarduzzi}
\ead{valentina.lenarduzzi@oulu.fi}

\address[OULU]{University of Oulu, Finland}
\address[UNIBZ]{Free University of Bozen-Bolzano, Italy}

\begin{abstract}
\textit{Background.} Technical debt (TD) has long been one of the key factors influencing the maintainability of software products. It represents technical compromises that sacrifice long-term software quality for potential short-term benefits.   \\
\textit{Objective.} This work is to collectively investigate the practitioners' opinions on the various perspectives of TD from a large collection of articles. We find the topics and latent details of each, where the sentiments of the detected opinions are also considered. \\
\textit{Method.} For such a purpose, we conducted a grey literature review on the articles systematically collected from three mainstream technology forums. Furthermore, we adopted natural language processing techniques like topic modeling and sentiment analysis to achieve a systematic and comprehensive understanding. However,  we adopted ChatGPT to support the topic interpretation. \\
\textit{Results.} In this study, 2,213 forum posts and articles were collected, with eight main topics and 43 sub-topics identified. For each topic, we obtained the practitioners' collective positive and negative opinions. \\
\textit{Conclusion.} We identified 8 major topics in TD related to software development. Identified challenges by practitioners include unclear roles and a lack of engagement. On the other hand, active management supports collaboration and mitigates the impact of TD on the source code. 

\end{abstract}


\end{frontmatter}

\section{Introduction}

Imagine purchasing a high-end product today, enjoying all its benefits immediately, but deferring the payment to a future date. While convenient in the short term, this financial strategy often incurs interest and can lead to a significant burden over time. This common scenario nowadays is called financial debt \cite{allman2012managing}.

Similarly, in software development, stashing technical debt (TD) allows teams to deliver features quickly but at the cost of future agility and efficiency \cite{allman2012managing, lenarduzzi2019technical}. Just as financial debt can grow and compound, so too can TD, transforming initial gains into long-term liabilities that can cripple a project's progress and sustainability. 

Therefore, TD is a commonly adopted metaphor in software engineering that represents the implied cost of future reworking resulting from the earlier choices of easy but limited solutions. Since Cunningham first brought up the concept \cite{cunningham1992wycash}, it has received increasing attention in the last decade. 

Though it is often considered an issue to be addressed in software development and maintenance practice, it is also well-acknowledged that TD can be both good and bad. One of the main advantages of incurring TD is the reduced maintenance time and cost in a given short time, which can help critical in-time deliveries and speed up the initial development \cite{guo2016exploring}. Especially when such technical debt, different from financial debt, might not require interest to be paid, which makes it nearly pure gain. Besides such a benefit in development time saving, TD can also help companies save startup capital and decrease risks \cite{besker2018embracing}. Furthermore, it is also beneficial that practitioners introduce TD to focus resources and efforts on achieving time-critical business goals and enable proper prioritization \cite{de2021business}. 

On the other hand, the direct and indirect issues caused by the TD are well-recognized by many studies, e.g., increased maintenance cost \cite{seaman2011measuring}, reduced code quality and system reliability \cite{bavota2016large}, slow feature development \cite{martini2015investigating}, decreased team morale \cite{besker2020influence}, and so on. 

Nonetheless, past studies focus on the symptom and not on the cause. Our work departs from the previous empirical studies based on code and metrics and aims to analyze practitioners' opinions on technical debt. To the best of our knowledge, none of the existing literature investigated practitioners' opinions on TD surveying the grey literature. Our main contributions are as follows:
\begin{itemize}
    \item large-scale opinion mining on TD by surveying grey literature.
    \item use of large language model (LLM) for data extraction and synthesis.
    \item identification of eight major topics, and 43 sub-topics for future research and practitioners' reference.
\end{itemize}

Future research efforts should focus on the developer's issues with TD while consolidating and improving the good practices. Moreover, research efforts should also focus on AI-driven TD detection and management, as well as an analysis of the effect of TD on current software development methodologies such as DevOps.

\textbf{Paper Structure}.  Section~\ref{sec:EmpiricalStudy} describes the design study we conducted. Section~\ref{sec:Results} describes the obtained results, while Section~\ref{sec:Discussion} discusses them. Section~\ref{sec:TV} highlights the threats to validity of this work. Section~\ref{sec:RW}  describes the related work, while Section~\ref{sec:Conclusion} draws the conclusion and future works.

\section{Study Design}
\label{sec:EmpiricalStudy}
This section describes our study's goals, research questions, and the methodology for answering them. Generally, to analyze the collective opinions of software practitioners on TD, we first collect a representative amount of articles from popular technology forums. With such text data, by adopting Natural Language Processing (NLP) techniques, e.g., topic modeling and sentiment analysis, we extract the main topics discussed on TD and the positive and negative opinions regarding each topic. In particular, we use ChatGPT to support the effective summarization of each topic based on the according keywords.
Our empirical study follows the guidelines defined by Wohlin et al.~\cite{DBLP:books/daglib/0029933}.

\subsection{Goal and Research Questions}

The \textit{goal} of our study is to conduct a large-scale survey of grey literature with the \emph{purpose} of identifying practitioners' most discussed aspects of TD and their sentiment connected to it. Our \textit{perspective} is of the practitioners who face the daunting task of paying or accumulating TD. The \textit{context} is forum posts and online articles discussing aspects of TD. 

Based on our goal, we defined the following three Research Questions.

\begin{boxC}
		\textbf{RQ$_1$.} \emph{What aspects are commonly discussed when practitioners talk about TD?}
\end{boxC}	 

Practitioners in the past weren't always aware of TD concepts and their impacts on their source code \cite{seaman2011measuring}. Nowadays, with the increasing focus on software quality both from academia and industry \cite{lenarduzzi2019technical}, practitioners have come to be knowledgeable about their decisions. Albeit TD sometimes is opaque to them, the proportion of ``conscious'' TD, or so-called self-admitted TD, is non-negligible \cite{fucci2021waiting,maldonado2015detecting}. Hence, practitioners have opinions on it and, based on their needs and constraints, they find themselves bound to accumulate TD to meet expectations and deadlines \cite{spinola2013investigating,kozanidis2022asking}

Therefore, in RQ$_1$, we aim to find the topics commonly discussed in the articles and discussion threads posted by practitioners regarding TD to discover practitioners' most focused aspects of TD. Identifying only the main topics does not contribute deeply to the state of the art nor allows for extracting comprehensive conclusions. Hence, we ask:

\begin{boxC}
		\textbf{RQ$_2$.} \emph{What are the issues about TD that the practitioners commonly complain about?}
\end{boxC}	 

Based on the main topics extracted in RQ$_1$, we further identify the sub-topics for each aspect of TD. In an optimal environment, practitioners have full knowledge and time to avoid stashing TD \cite{spinola2013investigating,ramavc2022prevalence}, but this cannot be further from reality \cite{ramavc2022prevalence}. Practioners have different sentiments tied to specific TDs \cite{fucci2021waiting} when self-admitting TD. Therefore, as suggested by Fucci et al. \cite{fucci2021waiting} use of sentiment analysis, in RQ$_2$, we analyze the content and the sentiment of the practitioner's discussions on TD, particularly focusing on their collectively negative opinions. Finally, we want to focus on the positive aspects discussed by the practitioners. Hence we ask:

\begin{boxC}
		\textbf{RQ$_3$.} \emph{What benefits can (paying) TD bring?}
\end{boxC}	 

Like in finance, paying a debt can have positive impacts \cite{ramavc2022prevalence}. Therefore, in RQ$_3$, we are interested in investigating the positive opinions regarding TD, possibly including the benefits of repaying TD and effective strategies to solve TD issues.

\subsection{Data and Processing}

Here, we introduce how and from where the data is collected and processed for analysis.

\subsubsection{Keyword selection}

To collect all the relevant texts from the target sources, we use "\textit{TD}" as the search query, which is straightforward but effective. First of all, this study must include the initial text selection. Therefore, we do not add any other keywords to limit the results. Meanwhile, we also aim to avoid being overly inclusive with an unnecessary amount of irrelevant articles. For example, we did not attach the abbreviation of ``TD'', \textit{TD}, as another keyword using ``OR'' because texts on HTML programming can also be included (\textit{<td>} is the tag defines a standard data cell in an HTML table). It is nearly universal that any articles on TD that use the abbreviation mention the full term at least once. Furthermore, we didn't consider the query ``\textit{technical} OR \textit{debt}'' for the same reason. 


\subsubsection{Data Source selection}
\begin{sloppypar}
To obtain the opinions of practitioners, we extract such content of the discussion and idea-sharing articles from three popular technology forums, i.e., \textit{StackOverflow}, \textit{Medium} and \textit{DZone}. StackOverflow is the largest forum of technology-related questions and answers (Q\&As) for developers and tech enthusiasts. Compared to StackOverflow, DZone is also one of the world’s largest online communities for developers but focuses more on new tech trends, e.g., DevOps, AI, big data, Microservices, etc. Furthermore, like DZone, Medium is a well-known technology forum that provides tutorials and review articles. Comparatively, articles on Medium are more common-reader-friendly and written in a non-technical style. These three platforms are the largest tech communities that can be considered to cover a representative school of opinions described in different styles. 
\end{sloppypar}

\begin{figure}[ht]
    \centering
    \includegraphics[width=.4\textwidth]{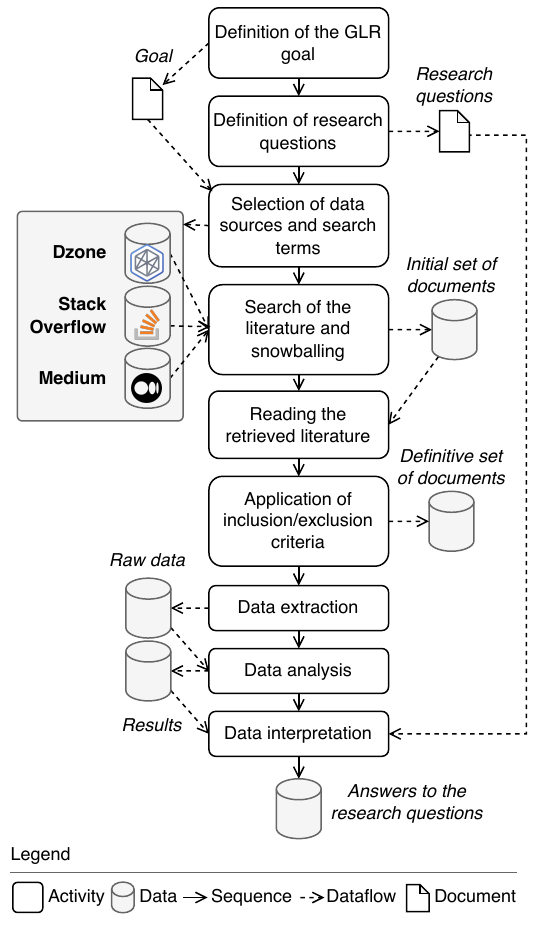}
    \caption{Overview of the followed process (adapted from \cite{janes2023open})}
    \label{fig:mlrprocess}
\end{figure}

\subsubsection{Data Extraction}

Due to the different availability of APIs and different crawling policies of these three platforms, we applied different data crawling strategies for each platform accordingly:

\begin{itemize}

    \item \textit{StackOverflow}. We applied API-based content crawling using the StackExchange API to retrieve the questions and answers regarding each selected tracing tool. Specifically, we used the advanced search API\footnote{\url{https://api.stackexchange.com/docs/advanced-search}} to extract all the questions that contain the name of the tool in either the title or the body of the question, together with the according answers. Please note that due to the API's daily query limitation, the \textit{pagesize} parameter was set to the maximum (i.e., 100 results shown per query) to minimize the crawling time. 
    
    \item \begin{sloppypar}\textit{Medium}: Though Medium's policy does not allow direct crawling from the forum, instead, it provides \textit{archive} for the articles assigned with tags. All the historical articles with a given tag $T$ can be found at \textit{https://medium.com/tag/T/archive}. Therefore, by using tools like BeautifulSoup\footnote{\url{https://www.crummy.com/software/BeautifulSoup/}}, we could obtain all the corresponding articles by traversing all the time-stamp URLs.   
    \end{sloppypar}
    
    \item \textit{DZone}: We adopted a hybrid crawling approach combining manual search and BeautifulSoup. Unlike Medium, Dzone allows crawling with BeautifulSoup within each article but not within the search results list. Hence, we conducted a hybrid crawling strategy by manually collecting all article URLs for each tool and then automatically crawling the article content for each URL using BeautifulSoup.
\end{itemize}

\subsection{Data Analysis}

\begin{figure}[!ht]
\centering
  \includegraphics[width=.4\textwidth]{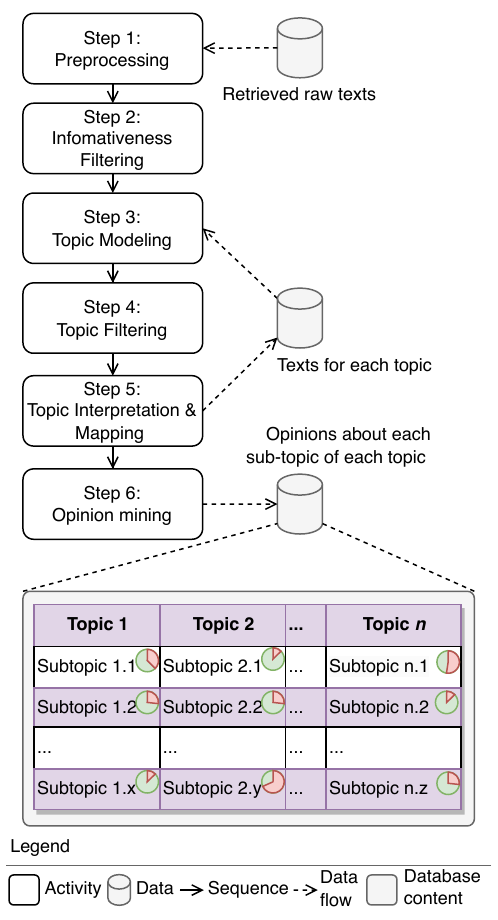}
  \caption{Data Analysis process (adapted from \cite{janes2023open})}
  \label{fig:nlpframe}
\end{figure}

By following the steps shown in Figure \ref{fig:nlpframe}, we can extract the topics commonly discussed by the practitioners and their collective opinions on each topic. The process contains five steps, which are described below.



\begin{itemize}
    \item \begin{sloppypar}\textit{Step 1: Preprocessing}: This step pre-processes the raw text data and prepares them for further analysis. First, we divide texts from the dataset into sentence-level instances since each text can contain multiple topics and various sentiments. Second, we build the bigram and trigram models, which means we identify the common phrases (e.g., \textit{New York} instead of \textit{new} and \textit{york}). Subsequently, a series of text processing activities are required for each sentence, including transforming text into lower cases, removing non-alpha-numeric symbols, screening stop-words, eliminating extra white spaces, and lemmatization.
    \end{sloppypar}
    
    \item \textit{Step 2: Informativeness Filtering}: This step filters out the noninformative sentences with a trained text classifier. By doing so, we shall identify the sentences that contain useful information and screen out those not relevant. Aiming to answer the research questions, the informative sentences shall contain explicit discussion regarding TD. For example, ``\textit{In the context of software development, TD is a shortcut taken by the developer to save time and finish a given project faster than they would otherwise be able to.}'' is informative by describing the concept and practice of TD in general; ``\textit{Do get in touch with us at hello@xxx.com to inquire about how we can help you build your product.}'' is non-informative and should be filtered out. 
    
    \item \textit{Step 3: Topic Modeling}: Herein, we detect the main topics of the informative sentences identified from the previous step. We shall determine the main aspects the articles discuss using topic modeling techniques. Selecting a topic model that fits the data is recommended to obtain topics that more accurately reflect the text data. For example, as the texts are tokenized to sentence level, a short-text topic modeling method, e.g., Gibbs Sampling Dirichlet Multinomial Mixture (GSDMM), is more effective than traditional Latent Dirichlet Allocation (LDA). Furthermore, if possible, domain-specific topic models can also be adopted.
    
    \item \textit{Step 4: Topic Filtering}: In this step, by using the first topic model built with the informative texts in the previous step, we could map each sentence to the topic it belongs to. Supposedly, when the topics obtained are too general, we need to investigate further the topics discussed within each of the general topics, which can be seen as ``aspect". We can adopt the same topic modeling method as the previous step for this step. 
    
    \item \textit{Step 5: Opinion Mining}: Finally, using opinion mining techniques on the texts, we can also know the collective sentiment from the social media concerning each topic in each aspect. Therefore, the outcome shall provide a detailed reflection on the percentage of positiveness and negativeness for each topic regarding TD.
    
\end{itemize}

Especially regarding the interpretation of the topics from lists of keywords, we adapt the ChatGPT-driven techniques used in Janes et al.'s study \cite{janes2023open}. Using ChatGPT \cite{chatgpt} can support effectively interpreting and summarizing the topics based on these keywords. ChatGPT is an artificial intelligence chatbot developed by OpenAI that uses foundational large language models (LLMs) and is fine-tuned via supervised and reinforcement learning techniques. Though a newly emerging technique, ChatGPT has quickly gained overwhelmingly worldwide attention from industry and academia. Specifically, regarding the facilitation of text summarization, many early-stage studies have investigated the use of ChatGPT for such tasks \cite{yang2023exploring,luo2023chatgpt}.

Herein, for this topic extraction task, we adopted the newly released GPT-4 model\footnote{\url{https://openai.com/product/gpt-4}}. Compared to the legacy GPT-3.5 model, the new model has much higher reasoning capacity and conciseness. We initiate the topic extraction by entering a series of structured requests formatted as follows.

\begin{flushleft}
    \textbf{``Extract a short <MAIN TOPIC>-related topic for each of these lines of keywords.}
    \\\textbf{ }
    \\\textbf{<1st LINE OF KEYWORDS>}
    \\\textbf{...}
    \\\textbf{<$n$th LINE OF KEYWORDS>''}
\end{flushleft}

The replies received were also structured to correspond to the following requests.

\begin{flushleft}
    \textbf{``<1st TOPIC>: <Explanation>}
    \\\textbf{...}
    \\\textbf{<$n$th TOPIC>: <Explanation>''}
\end{flushleft}

The first and third authors compare the AI-extracted topics and the original keyword lists for validation.

\subsection{Verifiability and Replicability}

We have published the complete raw data in the replication package to allow our study to be replicated\footnote{\url{https://zenodo.org/records/13133205}}.
\section{Results} 
\label{sec:Results}

By following the steps of the methods provided in Section \ref{sec:EmpiricalStudy}, we present the results that can answer the proposed research questions.

\subsection{Step 1: Preprocessing.} We pre-processed the crawled textual data by retaining only the natural language sentences. Herein, we eliminated unnecessary content, such as the source code, URLs, publishing date, author info, etc. For Medium articles, we started eliminating the article heading that includes the publishing date and author info by splitting the string at the common last character of the part ``min read'' and selecting the later part. Subsequently, we used the sentence tokenizer from the Natural Language Toolkit (NLTK)\footnote{\url{http://www.nltk.org/}} to obtain the list of sentences from each article. As the tokenizer does not identify the source code or URLs, we eliminated them by selecting only the sentences ending with a period, an exclamation mark, or a question mark. 

First, we crawled data from social media, including StackOverflow (319 questions and 510 answers), Medium (1368 articles), and Dzone (121 posts). Furthermore, we clean the text data by using \textit{langdetect} Python package~\footnote{\url{https://pypi.org/project/langdetect/}} to filter the non-English texts, \textit{html2text} package~\footnote{\url{https://pypi.org/project/html2text/}} to filter the source code and HTML markdowns from each text and customized regular expression based methods to drop the source code. Thereafter, we have 318 StackOverflow questions, 509 StackOverflow answers, 1265 Medium articles, and 121 Dzone posts. Then, we tokenized the texts into sentences using the NLTK sentence tokenizer and obtained 3044 sentences in StackOverflow questions, 965 in StackOverflow answers, 49903 in Medium articles, and 6801 in Dzone posts.

\subsection{Step 2: Informativeness Filtering.} Herein, we identified the informative sentences using a Na{\"i}ve Bayes (NB) classifier or the Expectation Maximization for Na{\"i}ve Bayes (EMNB) classifier \cite{nigam2000text}. The study of Janes et al.~\cite{janes2023open} suggests the two options that provide very high performance. The selection shall be based on the accuracy comparison of these two classifiers with the obtained dataset. First, we manually labeled a sufficient number of training data, including 50\% informative sentences and 50\% half non-informative ones. The selection criteria for informative sentences are that the sentence must explicitly discuss TD. With increasing training and testing data, the two classifiers shall be trained and compared with the F1-score using a 5-fold cross-validation. 

\begin{figure}[ht]
\centering
  \includegraphics[width=0.48\textwidth]{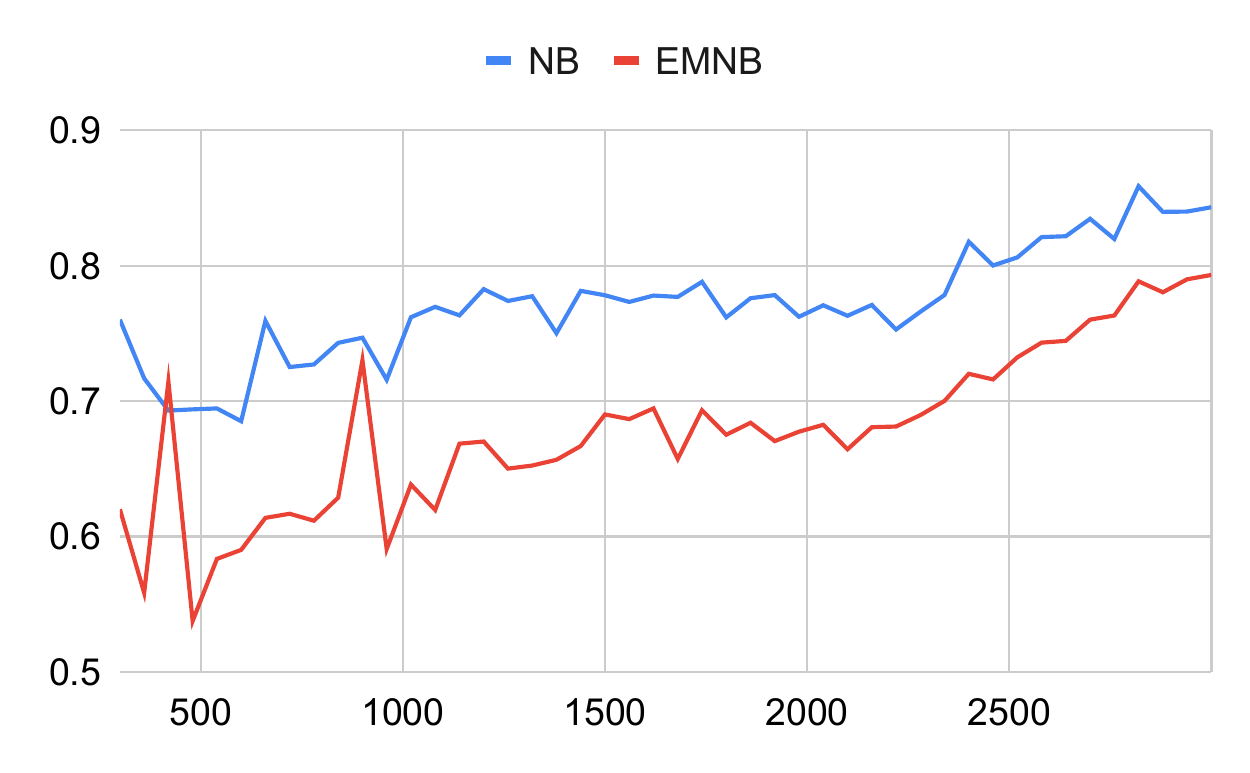}
  \caption{Testing informative text classifier accuracy}
  \label{fig:tclass_info}
\end{figure}

In this study, from the 60713 sentences obtained previously, we manually selected 3\,000 training data, including 1500 informative sentences and 1500 non-informative ones. To evaluate the performance of informativeness filtering, with a series of experiments, we compared the results of the NB algorithm and the EMNB algorithm with 3000 training data. We inspected the accuracy comparison of the two classifiers with different amounts of data starting from 200 to 3000 with an incremental step of 20. The test data ratio is set as default (0.25). Figure \ref{fig:tclass_info} shows that with the given training data, NB performs better than EMNB, with the accuracy reaching as high as 0.843. The classifier performs even better than the one used in \cite{janes2023open} (0.76). Thus, we adopted the NB classifier for filtering the informative sentences. Using the classifier trained by the 3000 training data, we obtained 36523 informative sentences. 

\subsection{Step 3: Topic Modeling.} To extract the topics from a large volume of textual data, we adopt the collapsed Gibbs Sampling Dirichlet Multinomial Mixture (GSDMM) \cite{yin2014dirichlet}. Initially, the textual data points (i.e., the sentences obtained previously) are randomly assigned to a pre-defined number of clusters with a set of critical document label information. Then the documents are traversed for several iterations, in each of which a cluster is re-assigned to each document according to a conditional distribution where these rules are followed: 1) choose a cluster with more documents, 2) choose a cluster where the documents therein have higher similarity. GSDMM has a much better performance on short texts compared with LDA \cite{weisser2023pseudo}; therefore, it is the better option for this study.

Herein, considering the effectiveness of the GSDMM model and the efficiency of the modeling process, we adopt the common default setting of the model with 30 iterations, $\alpha = 0.25$ and $\beta = 0.15$. Meanwhile, we set the upper bound topic number parameter $K = 30$. We consider Li et al.'s secondary study \cite{li2015systematic} summarized 24 TD-related notions when it would be highly unlikely the total topic number exceeds such a limit. Furthermore, we set the number of words for each topic to 30 to ease the topic interpretation.

\subsection{Step 4: Topic Filtering.} 

By training with the 36523 informative sentences, we obtained the GSDMM topic model with 20 topics. Therein, 35521 sentences (97.3\%) are associated with eight topics. Each of the other 12 topics contains no more than 2.3\% of the data; therefore, these topics are omitted. The number of sentences and extracted topics are reported in Table \ref{tab:maintopics}.

\begin{table}
  \centering
  \footnotesize
  \caption{The Main Topics on TD}
  \label{tab:maintopics}
  \begin{tabular}{l|p{6cm}|l}\hline
   \textbf{\#} & \textbf{Interpreted Topic} & \textbf{\#Sentences} \\ \hline
   1 & The True Cost of TD: Balancing Short-term with Long-term Pain & 2788 \\
   2 & Managing TD: a Team-Centric Approach & 10183 \\
   3 & Paying down TD in Agile Sprints & 1982 \\
   4 & Understanding the Business Impact of TD & 2480 \\
   5 & Measuring TD: Tools and Metrics & 1183 \\
   6 & TD in Product Development: A Necessary Evil? & 5063 \\
   7 & Balancing New Features and TD in Software Development & 7744 \\
   8 & Code Quality and TD: the Role of Testing & 4098 \\
   \hline
  \end{tabular}
\end{table}

Furthermore, we adopted the same process to train a separate GSDMM topic model for each subset of sentences associated with each main topic. We used the same parameters for the main topic model described above for such a process. In addition, to reduce the bias caused by the limited data volume, we selected the sub-topics associated with at least 100 sentences. We reached at least 3\% 

\subsection{Step 5: Topic Interpretation and Mapping.} 
Here are the eight popular topics commonly discussed by practitioners collectively on social media platforms. 

\vspace{2mm}
\textbf{[T1] The True Cost of TD: Balancing Short-term Gain with Long-term Pain.} One of the main topics discussed by the practitioners is the nearly unpredictable cost of TD as well as the trade-off of short-term gain and long-term pain. It is a common concern that accumulated TD can have insufferable financial implications in software projects. Although delayed mitigation of such TD can be costly for maintenance, the in-time quick solution is urgently needed. Furthermore, the sub-topics in this aspect are as follows. 

\begin{itemize}

    \item \textit{[T1.1] The Long-Term Consequences of Shortcuts in Software Design:} The potential long-term consequence of TD is commonly the main concern of practitioners, as the core of TD adoption is to meet the urgent deadline by taking necessary shortcuts. The unforeseeable consequences can be damaging. One article mentioned that \textit{among the consequences of TD, the loss of clients is by far the worst}. The more direct influence on the near future can also be \textit{with each new feature added, even more efforts are required for a developer to implement it}.
    

    \item \textit{[T1.2] Choosing Quick Solutions and Incurring TD:} The other side of the coin for incurring TD is the benefit of having quick solutions in time. The apparent benefit of such activity is saving time in the short term when additional rework and increased complexity can be inevitable to a certain extent. Acknowledgedly, in this way, issues e.g., \textit{developers become less and less likely to fix systemic issues, instead choosing to implement workarounds, layers of abstraction, and quick fixes}, also occur.
    

    \item \textit{[T1.3] Balancing Speed and Sustainability in TD Payment:} The impact of incurring TD on the sustainability in TD payment is another concern of the practitioners when the balance between speed and sustainability is needed by practitioners. For example, one opinion is that \textit{shipping imperfect code can be justified, but business leaders will ultimately shoulder the responsibility for paying down tech debt over time}. And \textit{slowly “paying down” tech debt while investing in writing new code} is the common strategy suggested.
    

    \item \textit{[T1.4] Financial Analogies in TD:} The financial analogies are also commonly adopted when the article introduces the concept of TD for a beginner audience. Meanwhile, much like debt in finance, the burden of TD can be unbearable at a certain point.


    \item \textit{[T1.5] High Cost of TD in Financial Terms:} When adopting financial debt as an analogy or metaphor, the practitioners often express their concern about the potential high cost in financial terms. Similar to the debt or loan concept in financial terms, the high interest can cost increasingly more on resources and development efforts. Such a cost can \textit{impact the business through disruption, revenue, fines, brand equity, among others"}.


    \item \textit{[T1.6] Short-Term Gains vs. Long-Term Costs in TD:} On the other hand, the short-term benefits are also critical to the TD concept, as it is the main reason to adopt it. Regarding the motivation behind such tendency, it is acknowledged by many that \textit{this is due to our present bias, the tendency of humans to prefer payoffs in the short term over those in the long term} and \textit{software engineers, in addition to having the same fundamental bias, normally have a deep desire to see their work shipped.}
    

    \item \textit{[T1.7] The Lifecycle of TD in Software Development:} The lifecycle of TD in software development is another concern in terms of the cost and balancing for TD. Practitioners acknowledge that \textit{throughout its lifecycle, a product-engineering team will need to constantly balance the need to move quickly and test ideas with building a maintainable and flexible codebase}. Meanwhile, it is also advocated that \textit{incorporating regular reduction of technical debt into the product lifecycle would create a more controlled and viable product that is more suitable to follow and serve future customer needs}.


\end{itemize}

\textbf{[T2] Managing TD: A Team-centric Approach.} Due to the high cost of accumulated TD and potential risk in the future, TD management is a critical aspect the practitioners need to proactively conduct through the project's lifecycle. Apparently, managing TD requires team effort where management skills are required towards the needs of effective communication and problem solving. The goal is to incorporate the in-time solution for fast delivery while preventing TD from hindering future development. Furthermore, the sub-topics in this aspect are as follows.


\begin{itemize}
    \item \textit{[T2.1] Solving TD through Team Collaboration:} The practitioners indicate team collaboration is one of the "best practices" to facilitate the effective mitigation of TD because \textit{finding and fixing the loop is a much harder problem and would require cross-team effort}. Especially, when coming to estimation and priority, one mentioned the experience of \textit{thinking of “priorities” is better, estimations are still necessary when it comes to team collaborations}, which emphasizes the importance of collaboration towards solving TD.
    
    
    \item \textit{[T2.2] Tools and Strategies for Reducing TD:} In order to properly manage TD, suitable and effective tools and strategies adoption are important. As mentioned by practitioners, \textit{You need to make tool choices that enable you to hire properly and allow you to scale appropriately, and give you the flexibility to do what you need to do as a company.} Meanwhile, practitioners also suggest that \textit{isolating messy code is a very effective strategy for dealing with unavoidable debts.} Furthermore, it is also acknowledged that \textit{having a clear understanding of what sort of technical debt you are going to run up will enable you to have a solid strategy for dealing with it}.
    

    \item \textit{[T2.3] Understanding and Managing TD as a Team:} In order to solve TD effectively, understanding TD in the first place is critical, which also requires team effort. Practitioners mentioned that it is common that \textit{most techies don’t really understand technical debt as a concept}. What is worse, even \textit{sometimes the general managers don't understand why it is important to invest time on technical debts}. Many practitioners emphasize that \textit{in order for the debt to be addressed within the agile process, we need the whole team to understand the state of the code and how it impacts delivery.}
    
    
    \item \textit{[T2.4] The Role in Team for Perception in TD:} Practitioners also need to pay attention to the different perceptions of TD from the different roles in a software team. As mentioned by practitioners, the understanding of TD differs between roles, such as general managers and technical staff. Other roles, such as architects and company leadership, are also mentioned to have an obligation to understand TD and its impact. 
    
    
    \item \textit{[T2.5] Team Involvement in Decision-Making on TD:} Based on the common and different understanding of TD across the team, together with the support of proper tools and strategies, the decision-making activity is also critical in terms of TD management, which also requires team involvement. Practitioners mention that \textit{communication gaps about the process leading to poor decision making, ... your technical debt load is increasing, you’ll continue to drown and reduce your team’s velocity.} On the other hand, it is also acknowledged that various decision-making will also lead to TD.
    
    
    \item \textit{[T2.6] The Role of Engineers in Managing TD:} Software engineers (i.e., developers) certainly play a crucial role in managing TD within software development, as they are mostly the direct inducer of TD as \textit{mostly, technical debt occurs because the developers write subpar code}. Other roles, like security engineer and data engineer, can also be influenced by TD management. For example, \textit{technical debt in your data pipeline will impact your organization in ways that will annoy stakeholders, make the working lives of analysts tedious, and frustrate data engineers.}
        

    \item \textit{[T2.7] Managing TD in Agile Teams:} Regarding team collaboration in managing TD,  many practitioners mentioned such practice and its importance in agile teams, as TD management is a continuous process for agile teams. One of the advantages is that \textit{.. clean code is linked to agile development and continuous delivery}. Meanwhile, \textit{adopting proper agile practices could help reduce and track technical debt.}    
    

\end{itemize}

\textbf{[T3] Paying Down TD in Agile Sprints}. As mentioned by many practitioners talking about managing TD with team collaboration, they consider "paying back" TD with continuous and agile practice is the right way. Therein, agile sprint planning is important in facilitating the management and reduction of TD by prioritizing, tracking and addressing them using project backlog. Furthermore, the sub-topics are as follows.


\begin{itemize}
    \item \textit{[T3.1] Allocating Resources for TD in Agile Sprints:} In general, many practitioners have mentioned that the allocation of time and resources of the project team is important when managing TD in agile sprints. For such a purpose, \textit{sprint planning meeting is an opportunity to decide how to use your allocated time} where the incoming sprint assignment towards solving TD can be decided collectively. By doing so, a clear allocation of effort for refactoring, fixing bugs, and improving code quality can be expected with the progression of well-planned sprints.

    
    \item \textit{[T3.2] Prioritizing TD in Sprint Planning: } When planning the sprints that integrate TD management activities, it is also important to prioritize the tasks together with the non-TD-related ones properly. Practitioners value activities that \textit{measure debt continuously in order to control the total cost of ownership of the application life-cycle and include debt measurement in project management and prioritization}. Meanwhile, one critical prerequisite towards efficient prioritization is the understanding of TD by stakeholders. 
    

    \item \textit{[T3.3] Addressing TD in Sprint Backlogs:} When the preparation towards solving TD is done, including resource allocation and prioritization mentioned above, TD can be addressed continuously in sprint backlogs, where these tasks should also be properly tracked and addressed alongside new features and bug fixes. The team should make sure TD is continuously addressed when it is also possible to \textit{devote an entire sprint to addressing a large amount of technical debt.}  
    
    
    \item \textit{[T3.4] Daily Focus on TD in Agile Teams:} From the project mentality perspective, it is also a good practice to have TD considered in the daily focus within agile teams. Practitioners mention that \textit{the difference between “eventually repaid” and “sustainable” is the amount of interest you pay on a daily basis while making changes to the code.} And practitioners acknowledge that it is important that \textit{we have to really consider eradicating technical debt as part of daily work}.
    
    
    \item \textit{[T3.5] Prioritizing TD in Sprint Backlogs:} Prioritization of TD-related tasks in sprint backlog is also mentioned by many. \textit{If the debt can be understood by the owner and prioritized in relation to other system requirements, then it may reasonably be managed on the product backlog.} This means assessing each item of TD, understanding its impact on the product and the user, and allocating appropriate resources to address it on time. By consistently including TD in the sprint backlog with elaborate estimation and prioritization, teams can ensure that it is paid down regularly.


\end{itemize}

\textbf{[T4] Understanding the Business Impact of TD}. Another key topic discussed by the practitioners is the business impact of TD and its understanding. The unchecked TD, when accumulated, can inevitably impact the development efficiency, which can eventually influence the software product quality and customer satisfaction. Furthermore, the sub-topics in this aspect are as follows.


\begin{itemize}
    \item \textit{[T4.1] The Risk Impact of TD for Company and Organization:} It is commonly acknowledged that TD will hurt the business of companies in the long run. \textit{With too much debt, you run a big risk of having to spend months rewriting your code, which is the worst situation to be in and is commonly the first step to eventually running out of money and shutting down the company.} Furthermore, at the company level, unchecked TD can result in the company's \textit{suffering costs from fixing the problem, providing credit monitoring services, reputational risk, or even facing regulatory action. Individuals are the ones fighting for months to repair their identity}.

    
    \item \textit{[T4.2] The Cost of TD on Business Development:} Reducing TD also means limiting its potential cost on the business level; however, \textit{the organization has little idea of how much technical debt it 'carries' in its code and paying tech debt is notoriously difficult to make visible to those setting business level priorities}. Meanwhile, \textit{many business managers can only see the value of the feature and not appreciate (unless you find ways to explain) the technical debt build-up, which will make everything cost more and take longer.} To track the cost of TD, \textit{Stepsize}\footnote{https://www.stepsize.com/technical-debt} is a tool recommended by the practitioners to help \textit{capture and track TD from their workflow so they can quantify its cost to the business and ultimately prioritize the most important debt.}
    
    
    \item \textit{[T4.3] TD Impact on Team Morale and Productivity:} Regarding the impact of TD, team morale and productivity is another critical concern of practitioners. It is commonly acknowledged that \textit{leaving technical debt unaddressed can result in many problems in your organization, such as low team morale} and \textit{tech debt will impact every corner of your business, from team morale to customer satisfaction.} Certainly, on the contrary, \textit{implementing a process for managing technical debt will positively influence engineering team morale and customer satisfaction.}
    

    \item \textit{[T4.4] Evaluating the Business Risks of TD:} It is critical to assess the risk of TD on the business level together with the estimation in development practices, as unchecked TD can pose significant risks to the organization as a whole. Commonly, as mentioned by the practitioners, \textit{the (main) risk (of TD) is that you will end up with an inferior product that also has more support and maintenance costs.} Some propose a potential good practice is to \textit{build a debt limit, which is essentially a measurement of risk and ability to make timely “payments.”} Meanwhile, another potential risk-reducing practice is \textit{quick feedback means a lot less risk than blocking feature development to slowly refactor the code in isolation}.



\end{itemize}

\textbf{[T5] Measuring TD: Tools and Metrics.} Tools and metrics that are adapted to measure TD properly are critical to the management of TD. The tools and metrics can quantify TD in a codebase, helping teams to track better, manage, and address code quality issues. Furthermore, the sub-topics in this aspect are as follows.


\begin{itemize}
    \item \textit{[T5.1] Tracking TD with Code Editors and Tools:} TD-tracing tools and plugins within code editors can largely facilitate the practitioners in terms of efficiency and code quality assurance. Code editors, e.g., VBSCode and JetBrains, offer extensions that can directly manage and track TD. For example, Stepsize is one of the tools for such purpose, and \textit{it’ll allow you to quickly report technical issues in your code without leaving your editor (see the VSCode or JetBrain ...)}. Such tools can provide various useful features besides TD tracking, e.g., link issues, prioritize workflow, and generate reports, etc.
    
    
    \item \textit{[T5.2] Measuring TD with SonarQube:} \textit{Types of Tools for Addressing Technical Debt Tools such as Stepsize, Sonarqube, and Klockwork are used to manage technical debt.}, as mentioned by practitioners. SonarQube (or SonarCloud) is one of the most popular tools that practitioners use to measure TD. However, practitioners also share their experiences (positive or negative) about SonarQube. For example, \textit{In Sonarqube versions prior to 5.5 there was the possibility to change the way that technical debt is calculated in order to take into account the complexity, but after 5.5 I can’t see how to change it}.

    
    \item \textit{[T5.3] Using Metrics to Measure and Track TD:} It is critical for the software project team to understand the metrics to measure TD in order to manage it. As mentioned by practitioners, \textit{... engineering teams need a clear understanding of the strategy and metrics for technical debt}. The often-mentioned metrics used for TD tracking and management include the number of bugs, code quality scores, and impact on the maintenance cost, etc.


    \item \textit{[T5.4] Analyzing TD Levels and Violations:} When measuring and tracking TD within the project, it is common to classify TD with priority and severity, where "level" is a term adopted correspondingly. Meanwhile, it is also true when practitioners have concern that \textit{it’s hard to calculate the exact level of Technical debt, you need to know the seriousness of the situation once it gets unsolvable}. On the other hand, the number of violations in terms of standards or defined rules can be considered useful variables for TD measurement.  
    

    \item \textit{[T5.5] Refactoring and Managing TD with SonarQube:} SonarQube is a popular tool adopted commonly by practitioners for measuring and managing TD. SonarQube can provide features that support efficient refactoring to reduce the identified TD by suggesting automatic refactoring for code violations.

\end{itemize}

\textbf{[T6] TD in Product Development: A Necessary Evil?} Debt in a broader sense, \textit{... technical or otherwise, is a necessary dimension of all business and product development practice}. Practitioners also discuss TD-related strategy and practice on the level of product development, where managing and "paying down" TD comes together with maintaining a product focus. Furthermore, the sub-topics in this aspect are as follows.


\begin{itemize}
    \item \textit{[T6.1] Balancing Quality and Speed in Product Development:} balancing quality and development speed is crucial when managing product development, especially when teams prioritize rapid delivery over quality. Many practitioners discuss the common phenomenon where \textit{at the very beginning of a project, developers tend to favor speed over quality, and it works... soon it becomes riskier, harder, and more time-consuming to add features}. It is a common claim from the practitioners that \textit{they want to grow their product while maintaining speed and quality and, therefore, low technical debt}. And product managers, engineers, and product owners must balance this with the need for quality and long-term value.


    \item \textit{[T6.2] The Trade-Offs of TD in Fast-Paced Markets:} 
    In fast-paced business environments, product teams often face the dilemma of delivering new features quickly to meet market demands or focusing on quality and incurring TD. It is common when \textit{the pressure to speed up the time to market has intensified how businesses deal with the risk of malfunctions, bugs, and overall defects throughout the customer experience}. Teams need to understand the implications of these choices and strive to find a balance that maintains quality without significantly slowing down development.
    
    
    \item \textit{[T6.3] TD from a Development Perspective:} From the development team's perspective, the balancing between product quality and fast delivery often occurs, which likely leads to many shortcuts taken and unattended. Critical decisions on taking shortcuts or conducting refactoring can

    
    \item \textit{[T6.4] Managing TD in Product Development Teams:} In product development, TD is a concern that requires attention from various stakeholders, e.g., data and security engineers, managers, and etc. Senior architects and project managers play a crucial role in understanding the implications of TD and guiding the team toward sustainable development practices. The balancing mentioned in the previous sub-topic is the practices executed often based on the collective expertise.  
    
    
\end{itemize}

\textbf{[T7] Balancing New Features and TD in Software Development.} New feature development is a critical part of software evolution and can provide potential new attractive designs or functions to the end users, which can maintain and enhance their satisfaction to the product. Therefore, managing TD requires the project teams to balance the need for adding new features and that for addressing TD. Such practice also requires strategic decisions about where to allocate time and resources. Furthermore, the sub-topics in this aspect are as follows.


\begin{itemize}
    
    \item \textit{[T7.1] Allocating Resources for New Features vs. TD Reduction:} As practitioners are aware that \textit{the problem is that every new feature adds debt or creates a placeholder for future debt}. Especially when the resources are limited, the team should be careful allocating time and effort to balance the new feature development and TD reduction. Inevitably, \textit{without any means of keeping on top of that debt, you can be sure that sometime in the future, you are going to be working on integrating a new feature and be continually blocked at every turn by brittle/inflexible technical decisions made in the past that now would require a Herculean effort to fix before even getting to your feature}.

    
    \item \textit{[T7.2] Managing Feature Development and TD in Agile Projects:} The challenge of balancing the addition of new features with the need to reduce the existing TD and proactively preventing new TD exist commonly in agile projects. It is common to face situation like \textit{...You are supposed to deliver a new feature by the end of the sprint; ... And sometime during the sprint, the development team realizes that if we want to implement it the correct way - ensuring engineering best practices, it will overspill and is not feasible to complete in the earlier planned timelines}. Such situation should be properly managed.
    
    
    \item \textit{[T7.3] Strategic Prioritization of Bug Fixes and Feature Development:} Effective management of TD involves prioritizing maintenance work, e.g., bug fixes, and code improvements while also working on new features. The project teams need to assess the long-term impact of TD and allocate sufficient time to address it alongside ongoing development tasks, where explicit prioritization is required the most.
    
    \item \textit{[T7.4] Integrating TD Management into Feature Development:} It is commonly acknowledged inevitably \textit{every new feature adds debt or creates a placeholder for future debt}, which suggests that it could be a good practice that TD management proactively integrated when any new feature is development. 
    
    \item \textit{[T7.5] Continuous Improvement of Code Quality Amidst Feature Expansion:} Maintaining a balance between feature development and technical debt reduction is crucial for code quality. It is commonly acknowledged that \textit{there is undoubtedly a trade-off between feature work and the work developers know is crucial for long-term code quality and stability.} Teams must focus on fixing bugs, conducting thorough testing, and implementing robust development practices to ensure long-term sustainability and functionality of the software.

\end{itemize}


\textbf{[T8] Code Quality and TD: The Role of Testing.} It is also commonly mentioned by the practitioners that the connection between code quality, TD reduction and testing in software projects. The potential best practices, automation and review process to reduce TD are considered important. Furthermore, the sub-topics in this aspect are as follows.


\begin{itemize}
    \item \textit{[T8.1] Data and Testing in Managing TD:} Regarding machine learning and data-driven application, managing TD requires also closer monitoring the data quality, model performance and so on. Regular testing and improvement of data pipelines are also crucial to address TD, especially when changes in data models can significantly impact feature performance and overall cost. Considering the situation where \textit{in order to improve the product, it’d be useful to consider performance optimization — to decrease response time under high load and to increase the maximum number of users that could work simultaneously without bugs}, it is also important to consider the management of TD therein.
    
    
    \item \textit{[T8.2] Automated Testing in Reducing TD:} Automated testing plays a vital role in managing TD, particularly in development environments with continuous integration and deployment. \textit{The low stakes environment in software is automated testing before deployment. A unit or integration test is a very basic yet powerful way for probing and sensing. Good architecture enables such tests.} By automating tests for new code and features, teams can quickly identify and address bugs, ensuring that the development process does not add to the TD. This approach is crucial for maintaining product quality and reducing manual testing efforts.
    
    \item \textit{[T8.3] Testing as a Tool for Quality Assurance in Software Development:} Testing is commonly considered an essential practice for maintaining code quality when it is also for minimizing TD. Testing can facilitate the practice of reducing the long-term cost associated with TD. As mentioned by practitioners, \textit{Having .. changes in the code, we should manage the testing side properly because we have to ensure quality and not cause any kind of regressions}. Furthermore, it is also acknowledged that \textit{as technology is getting more advance, it is now possible to test the quality of code using automated software tools.}
    
    
    \item \textit{[T8.4] Testing and Refactoring in Microservice Architectures:} There are practitioners discussing the testing and refactoring regarding microservice-based systems in terms of managing the TD therein. Similarly, in the context of microservices, testing is also the key to managing TD, particularly when changes in code and architecture are frequent. After all, \textit{if the development process and integration is properly structured, and the pooled microservices performs as expected, the overall TD risk of all layers of microservice structured debt improve}. Effective testing practice can also support refactoring process towards maintaining a clean, efficient and scalable codebase.

    
    \item \textit{[T8.5] Improving Code Quality through Testing and Review:} Towards code quality enhancement, testing and code review are also critical in terms of TD identification and management. Both activities can help the development team to monitor code performance, identify security issues, and improve overall software quality. \textit{Using an automated code review tool can help change the perception of code quality — especially for beginning developers}, who are not familiar with the concept of TD and necessary actions. Proactively managing TD and maintain high-quality standards are important and can be achieved through testing and code reviewing.

    
    \item \textit{[T8.6] The Impact of Testing on TD in Software Architecture:} Specially for software products with complex architecture, lack of testing can lead to an increase in TD or specifically architecture debt. Good testing practices, including design and documentation reviews, are essential for identifying architectural flaws and high-complexity areas in the code. These practices help in making informed changes and improvements, reducing the long-term TD associated with poor architectural decisions.

    
\end{itemize}

\subsection{Step 6: Opinion Mining.} 

Using the VADER method \cite{gilbert2014vader}, we can assess the sentiment of each informative sentence and the overall sentiment of each topic. To be noted, herein, we consider the percentage of positive, neutral, and negative sentences without considering the according sentiment strength. To determine the benefits and issues of each main topic, we compared each set of sentiment percentages to the average sentiment percentage of all sentences. 

\begin{figure}[!ht]
\centering
  \includegraphics[width=0.48\textwidth]{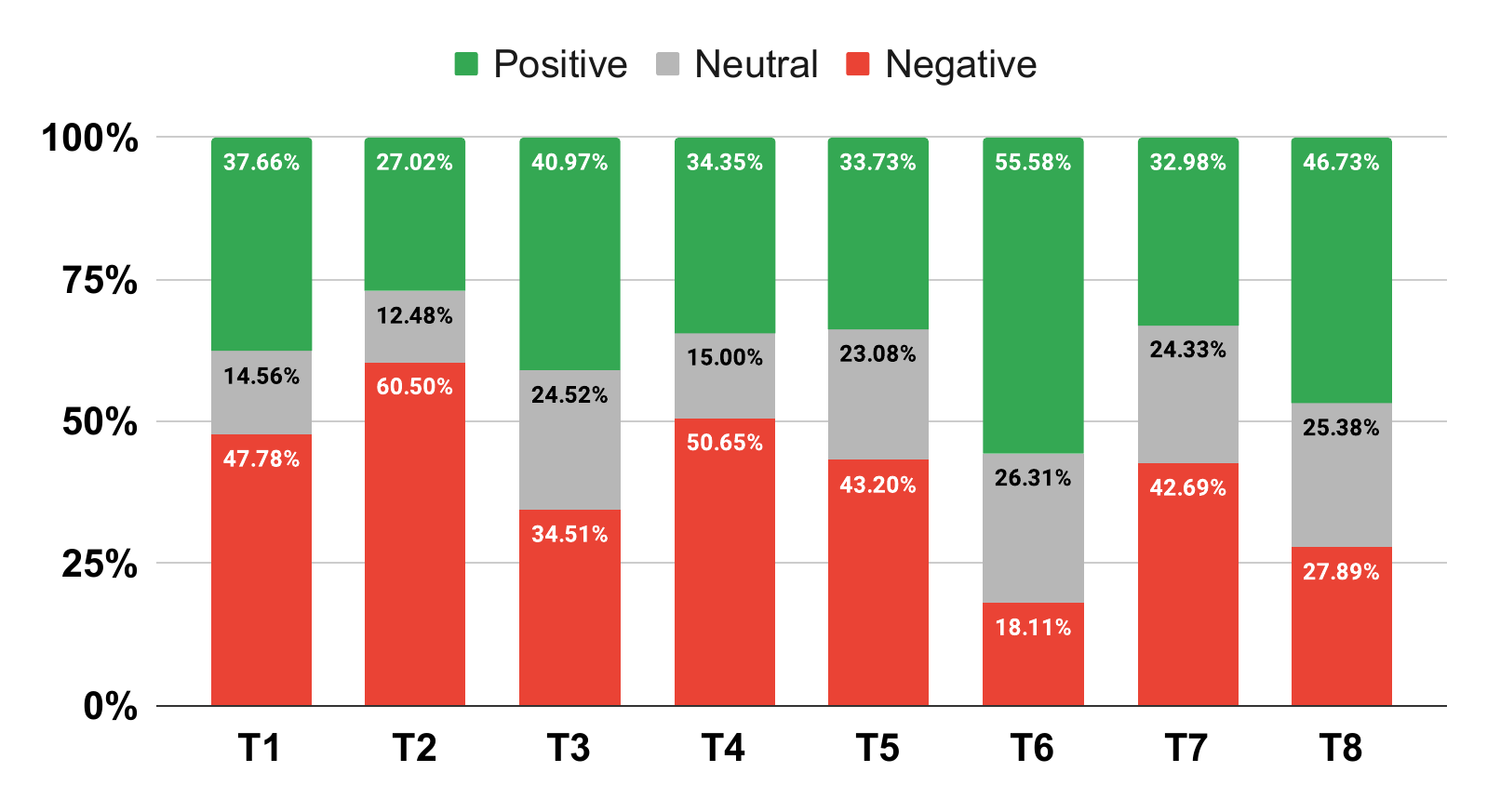}
  \caption{Topic Sentiment Summary}
  \label{fig:topicsent}
\end{figure}

The proportions of positive, neutral, and negative sentences for the main topics are shown in Figure \ref{fig:topicsent}. We can observe that in topics Managing TD (T2) and Understanding the Business Impact of TD (T4), the percentage of negative opinions (60.5\% and 50.6\%) are overwhelmingly more than the positive. Meanwhile, topics like The True Cost of TD (T1), Measuring TD (T5), and Dealing with TD in Agile Development (T7) also have considerably more negative opinions than positive ones. On the other hand, for the other three topics, the situation is reversed. Especially for the topic TD in Product Development: A Necessary Evil? (T6), the positive opinions are dominating.

\begin{figure*}[]
\centering
\begin{subfigure}[b]{0.48\textwidth}
    \includegraphics[width=\textwidth]{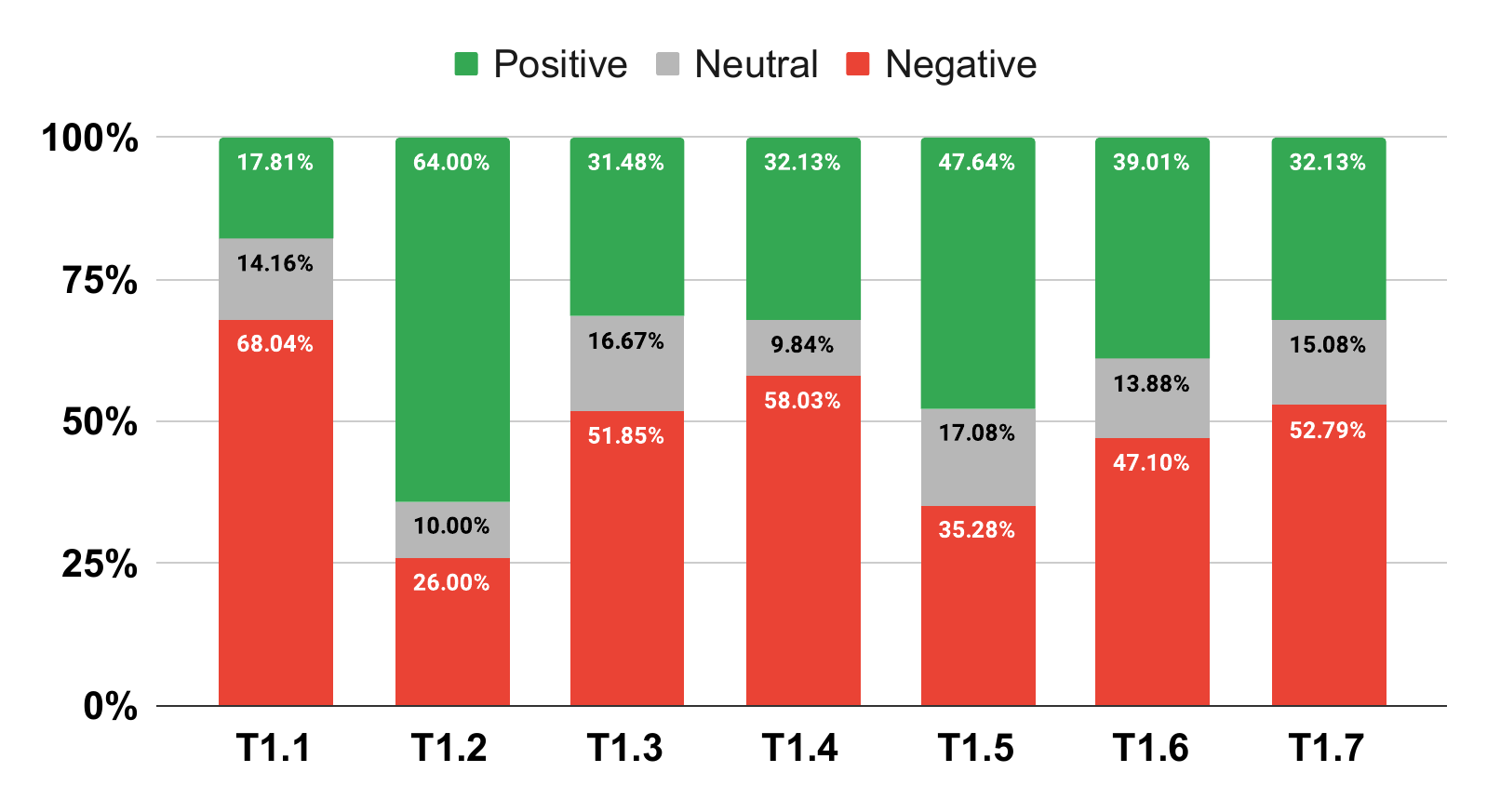}
    \caption{Sub-Topics Sentiments for Topic 1}\hspace{0pt}
\end{subfigure}
\begin{subfigure}[b]{0.48\textwidth}
    \includegraphics[width=\textwidth]{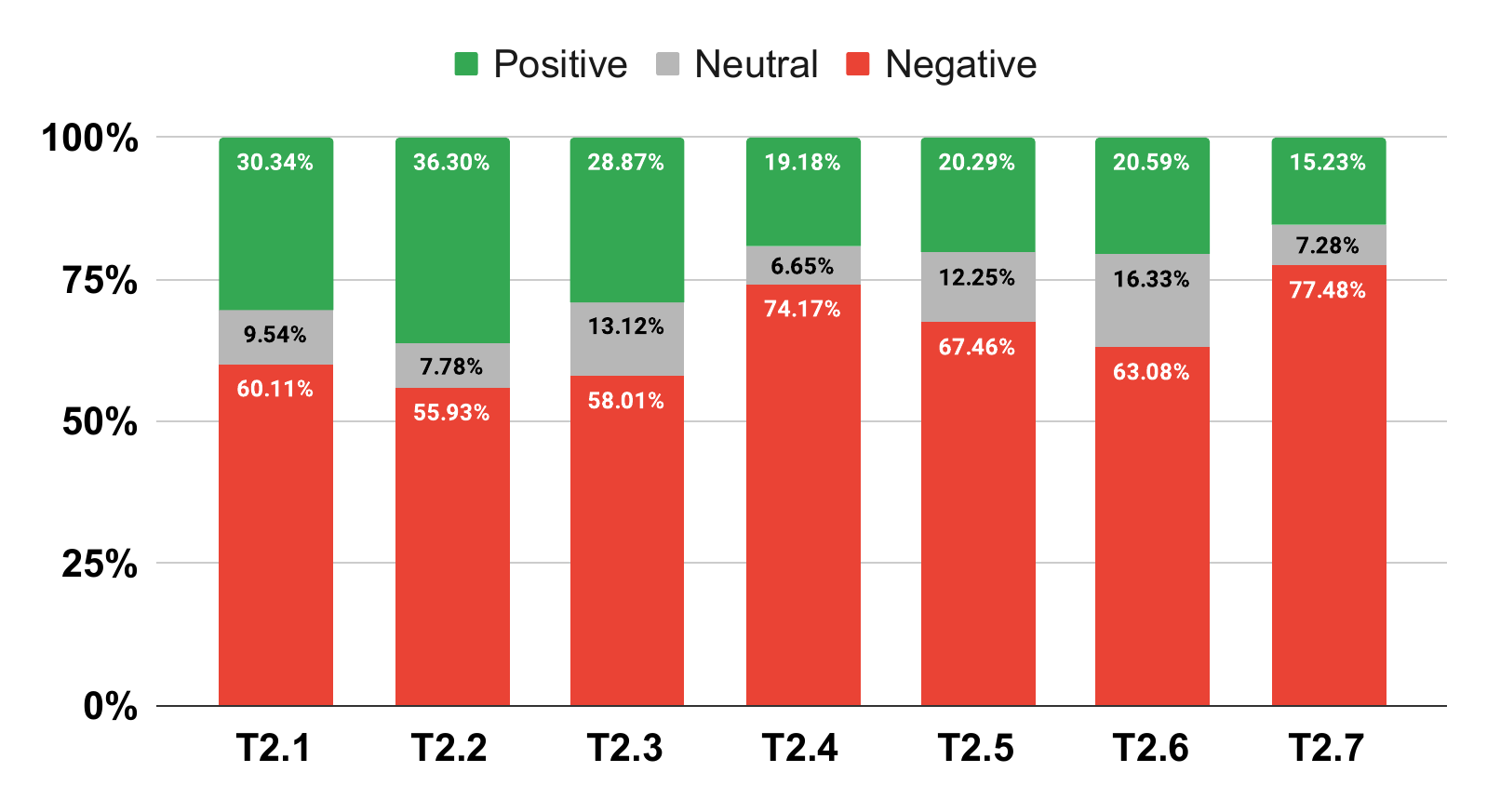}
    \caption{Sub-Topics Sentiments for Topic 2}\hspace{0pt}
\end{subfigure}
\medskip
\begin{subfigure}[b]{0.48\textwidth}
    \includegraphics[width=\textwidth]{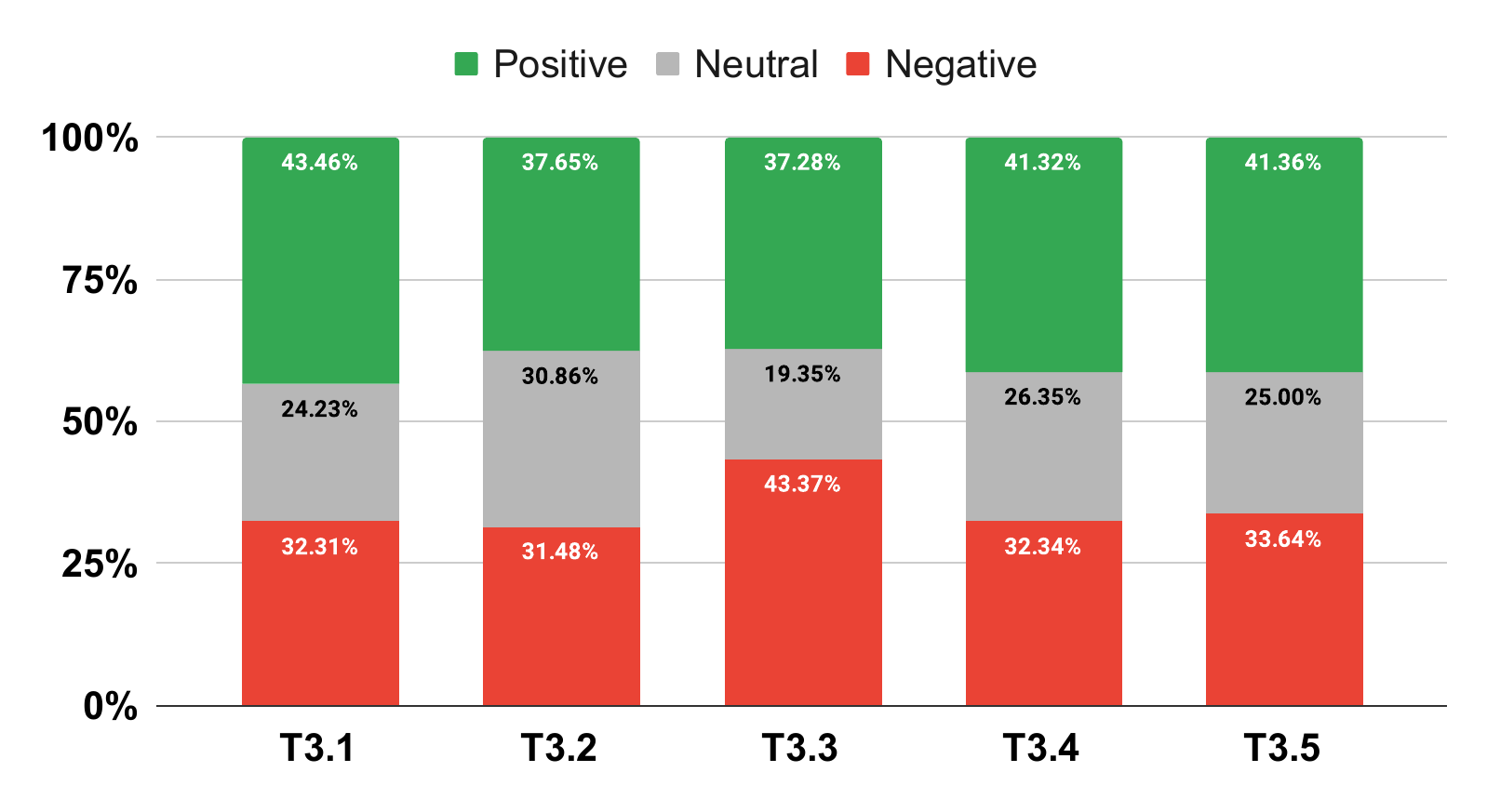}
    \caption{Sub-Topics Sentiments for Topic 3}\hspace{0pt}
\end{subfigure}
\begin{subfigure}[b]{0.48\textwidth}
    \includegraphics[width=\textwidth]{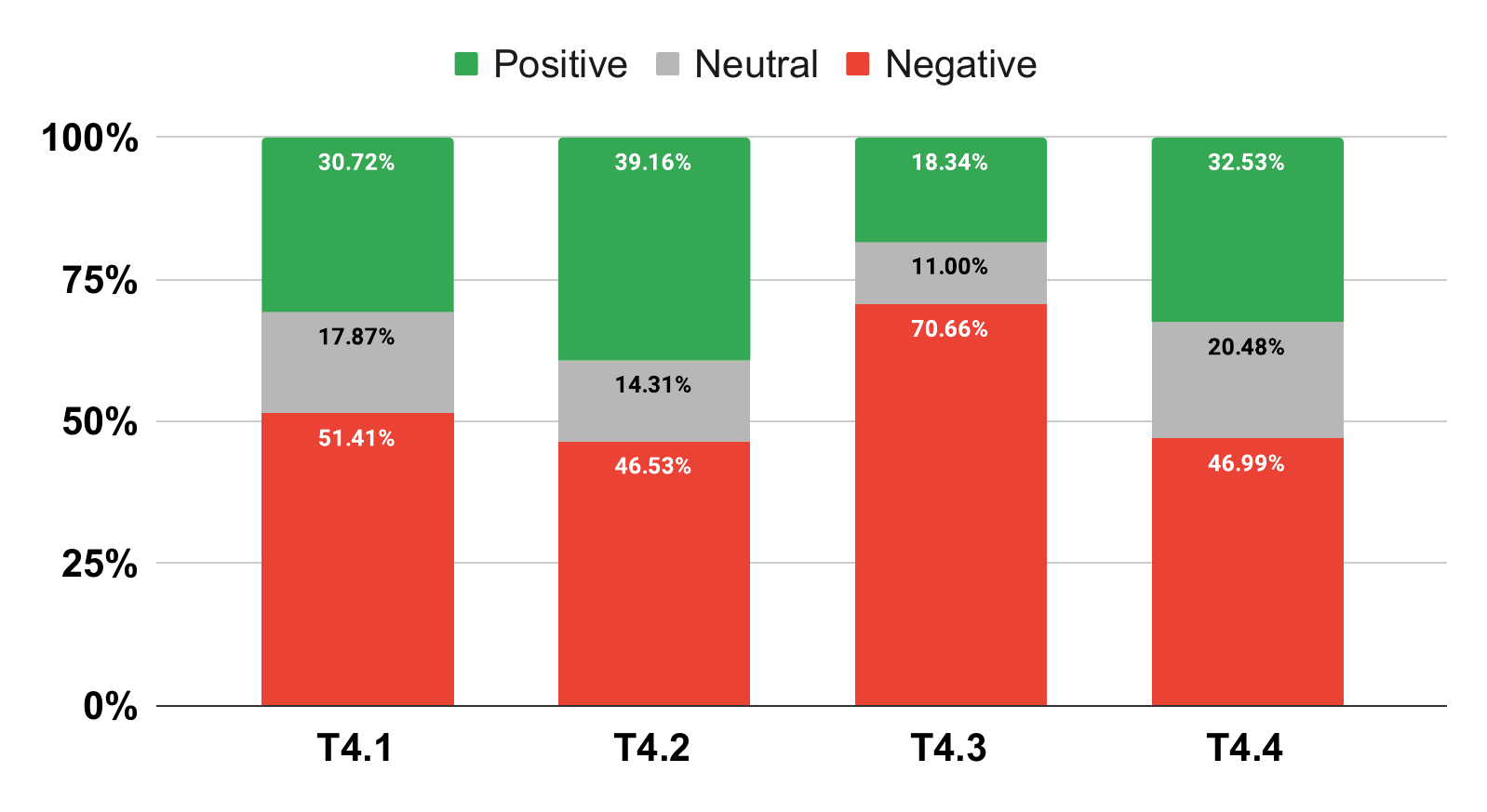}
    \caption{Sub-Topics Sentiments for Topic 4}\hspace{0pt}
\end{subfigure}
\medskip
\begin{subfigure}[b]{0.48\textwidth}
    \includegraphics[width=\textwidth]{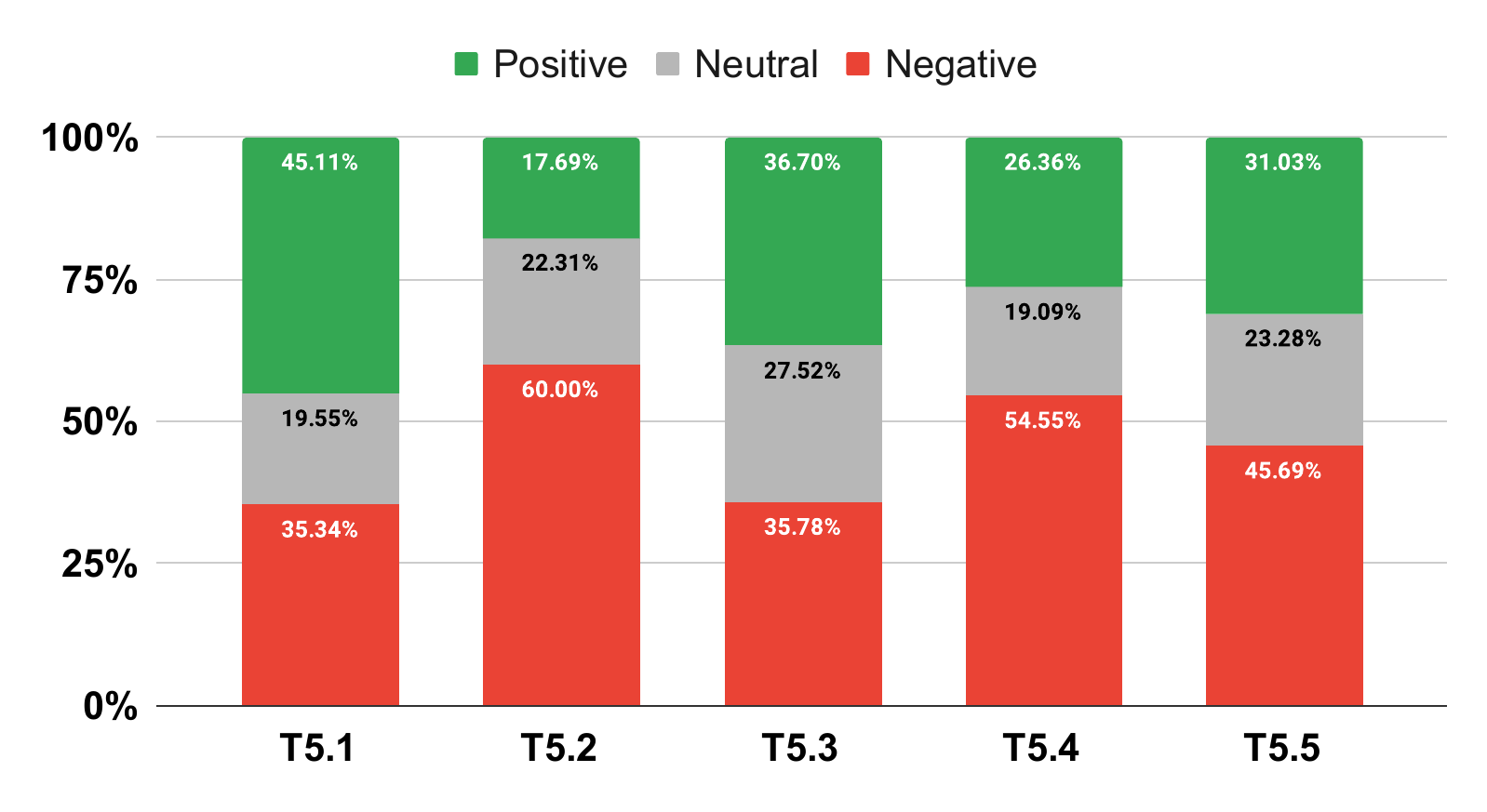}
    \caption{Sub-Topics Sentiments for Topic 5}\hspace{0pt}
\end{subfigure}
\begin{subfigure}[b]{0.48\textwidth}
    \includegraphics[width=\textwidth]{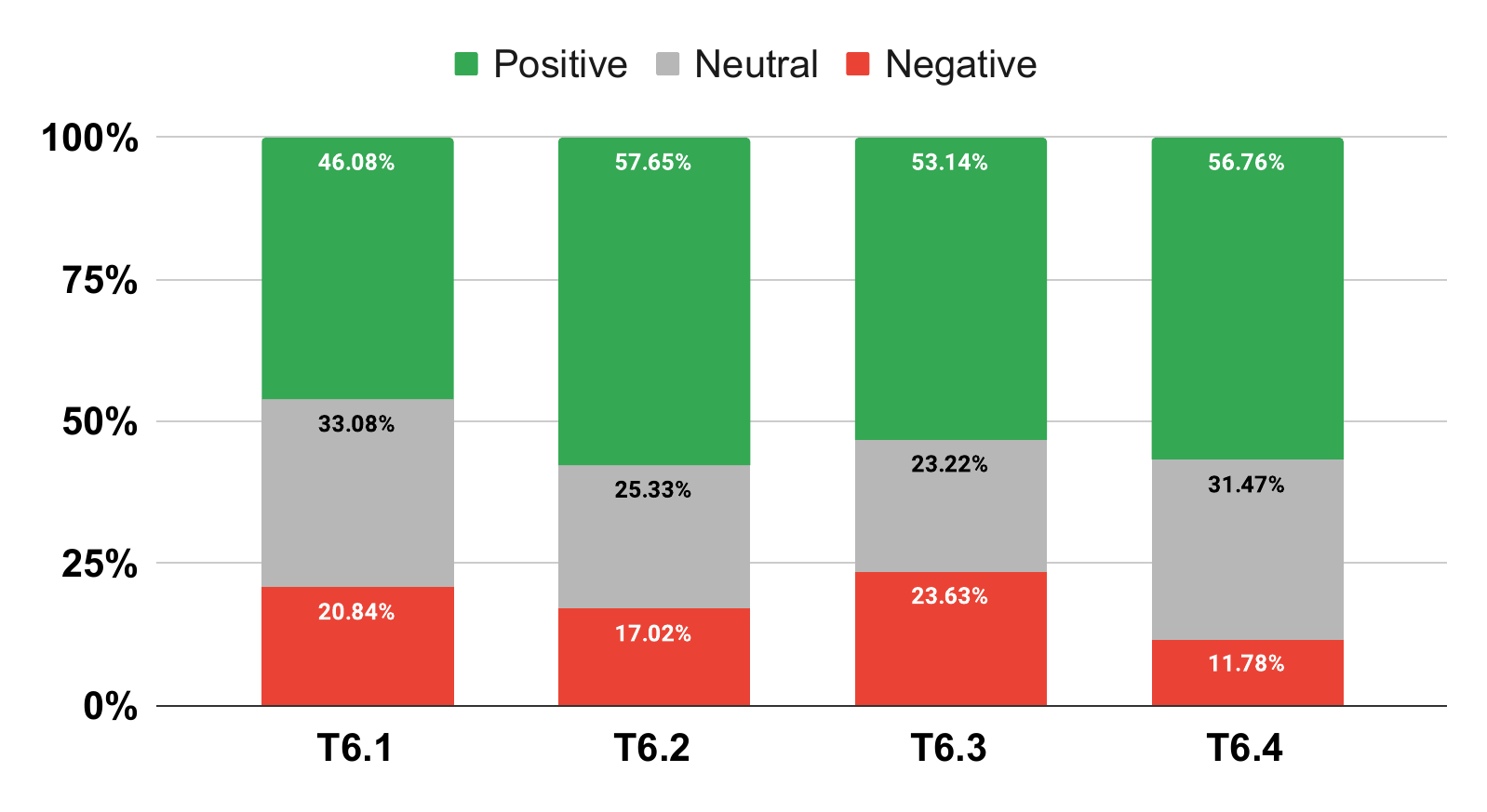}
    \caption{Sub-Topics Sentiments for Topic 6}\hspace{0pt}
\end{subfigure}
\medskip
\begin{subfigure}[b]{0.48\textwidth}
    \includegraphics[width=\textwidth]{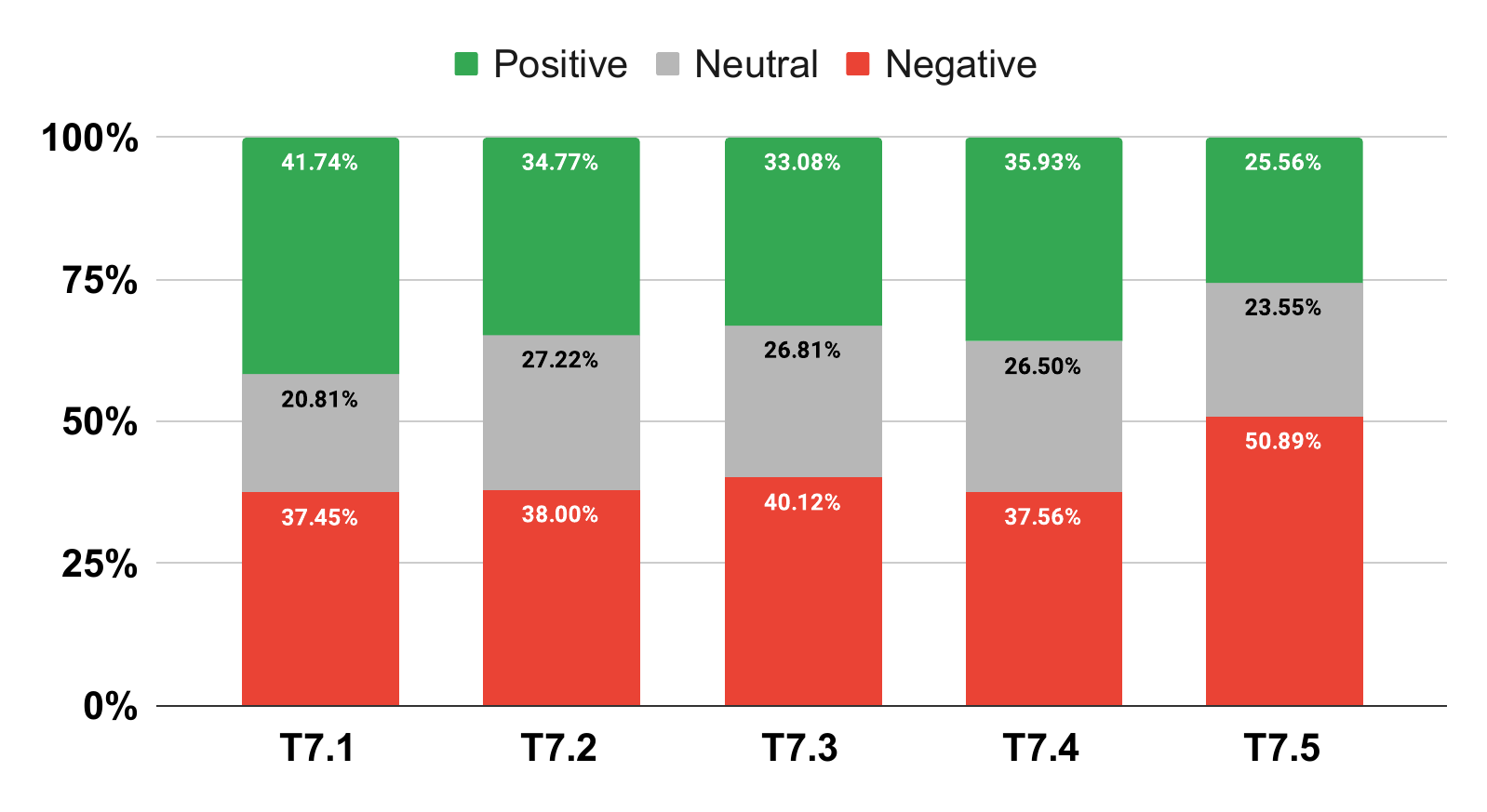}
    \caption{Sub-Topics Sentiments for Topic 7}\hspace{0pt}
\end{subfigure}
\begin{subfigure}[b]{0.48\textwidth}
    \includegraphics[width=\textwidth]{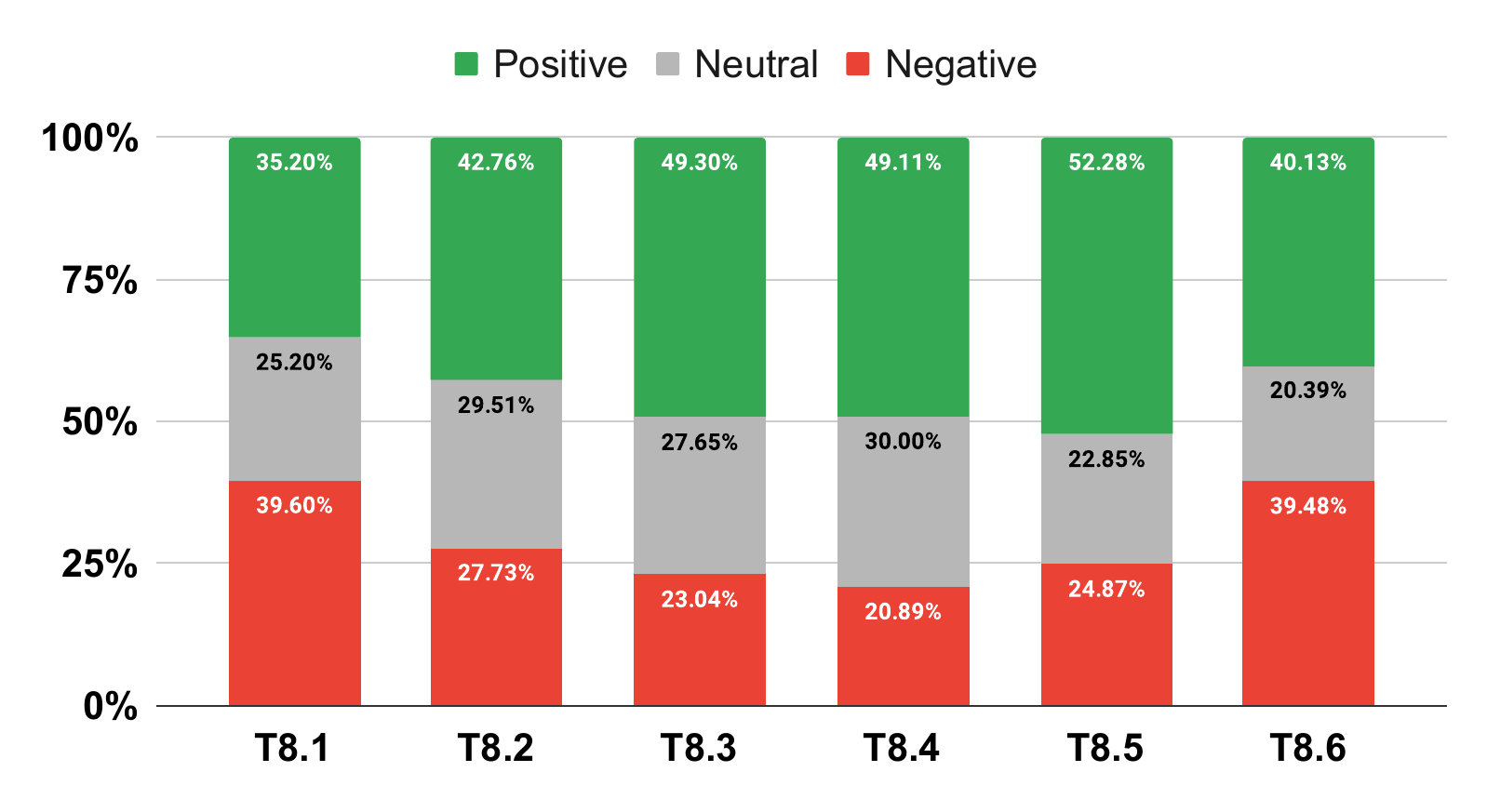}
    \caption{Sub-Topics Sentiments for Topic 8}\hspace{0pt}
\end{subfigure}
\caption{Sentiment Distribution for the Sub-Topics}
\label{fig:subtopicsent}
\end{figure*}

Furthermore, we also summarize the sentiment of the extracted sub-topics of each main topic, shown in Figure \ref{fig:subtopicsent}. By comparing the sentiment percentage between each sub-topic and that of its main topic, we determined each sub-topic for each main topic to be either a benefit, an issue, or a neutral opinion according to the following criteria. These criteria were also adopted by Janes et al. \cite{janes2023open}.

\begin{itemize}
    \item If the percentage of positive sentences is higher than average and the percentage of negative ones lower than average, the sub-topic is considered as a benefit of TD or a positive impact of handling TD.
    \item If the percentage of positive sentences is lower than average and the percentage of negative ones higher than average, the topic is considered an issue of TD.
    \item For any other circumstances, the topic is considered disputed.
\end{itemize}

\begin{table}[!ht]
    \setlength{\tabcolsep}{4pt}
    \centering
    \footnotesize   
    \caption{The Positive and Negative Opinions in Sub-topics}
    \label{tab:posnegsub}
    \begin{tabular}{L{2cm}|L{6cm}}
        \hline 
        \textbf{Topic} & \textbf{Sub-Topics} \\
        \hline
        
        [T1] The True Cost of TD & \negtd{[T1.1] The Long-Term Consequences of Shortcuts in Software Design;} \postd{[T1.2] Choosing Quick Solutions and Incurring TD;} \negtd{[T1.3] Balancing Speed and Sustainability in TD Payment;} \negtd{[T1.4] Financial Analogies in TD;} \postd{[T1.5] High Cost of TD in Financial Terms;} \postd{[T1.6] Short-Term Gains vs. Long-Term Costs in TD;} \negtd{[T1.7] The Lifecycle of TD in Software Development;} \\\hline
        
        [T2] Managing TD & \postd{[T2.1] Solving TD through Team Collaboration;} \postd{[T2.2] Tools and Strategies for Reducing TD;} \postd{[T2.3] Understanding and Managing TD as a Team;} \negtd{[T2.4] The Role in Team for Perception in TD;} \negtd{[T2.5] Team Involvement in Decision-Making on TD;} \negtd{[T2.6] The Role of Engineers in Managing TD;} \negtd{[T2.7] Managing TD in Agile Teams;} \\\hline
        
        [T3] Paying Down TD in Agile Sprints & \postd{[T3.1] Allocating Resources for TD in Agile Sprints;} [T3.2] Prioritizing TD in Sprint Planning; \negtd{[T3.3] Addressing TD in Sprint Backlogs;} \postd{[T3.4] Daily Focus on TD in Agile Teams;} \postd{[T3.5] Prioritizing TD in Sprint Backlogs;} \\\hline
        
        [T4] Understanding the Business Impact of TD & \negtd{[T4.1] The Risk Impact of TD for Company and Organization;} \postd{[T4.2] The Cost of TD on Business Development;} \negtd{[T4.3] TD Impact on Team Morale and Productivity;} [T4.4] Evaluating the Business Risks of TD; \\\hline
        
        [T5] Measuring TD: Tools and Metrics & \postd{[T5.1] Tracking TD with Code Editors and Tools;} \negtd{[T5.2] Measuring TD with SonarQube;} \postd{[T5.3] Using Metrics to Measure and Track TD;} \negtd{[T5.4] Analyzing TD Levels and Violations;} \negtd{[T5.5] Refactoring and Managing TD with SonarQube;} \\\hline
        
        [T6] TD in Product Development & \negtd{[T6.1] Balancing Quality and Speed in Product Development;} \postd{[T6.2] The Trade-Offs of TD in Fast-Paced Markets;} \negtd{[T6.3] TD from a Development Perspective;} \postd{[T6.4] Managing TD in Product Development Teams;} \\\hline

        [T7] Balancing New Features and TD in Software Development & \postd{[T7.1] Allocating Resources for New Features vs. TD Reduction;} \postd{[T7.2] Managing Feature Development and TD in Agile Projects:;} \postd{[T7.3] Strategic Prioritization of Bug Fixes and Feature Development;} \postd{[T7.4] Integrating TD Management into Feature Development;} \negtd{[T7.5] Continuous Improvement of Code Quality Amidst Feature Expansion;} \\\hline
        
        
        [T8] Code Quality and TD & \negtd{[T8.1] Data Metrics and Testing in Managing TD;} [T8.2] Automated Testing in Reducing TD; \postd{[T8.3] Testing as a Tool for Quality Assurance in Software Development;} \postd{[T8.4] Testing and Refactoring in Microservice Architectures;} \postd{[T8.5] Improving Code Quality through Testing and Review;} \negtd{[T8.6] The Impact of Testing on TD in Software Architecture;} \\
        \hline 
    \end{tabular}
\end{table}

Based on the criteria given above, we can identify the positive and negative sub-topics within each main topic theme. Shown in Table \ref{fig:subtopicsent}, the positive sub-topics (i.e., benefits) are marked in green color, while negative ones (i.e., issues) are marked red. 

Regarding the topic \textbf{[T1] The true cost of TD: Balancing Short-term Gain with Long-term Pain}, it is inevitable that practitioners have overall negative opinions about the long-term consequences caused by TD (T1.1) in a landslide. Meanwhile, the sentiment is overwhelmingly positive when discussing TD as a quick solution (T1.2). To a certain extent, it approves that the positive perspective of TD as a quick solution is commonly recognized. Furthermore, when discussing the balance between the speed for paying back TD and the sustainability between paying back and quick solutions (T1.3) and using the financial analogies as the metaphor (T1.4), the collective opinion is surprisingly negative. On the other hand, surprisingly, practitioners have positive opinions on the high cost of TD in financial terms (T1.5).
Similarly, the comparison between the short-term gain and long-term cost in TD (T1.6) and the relevant discussion is also positively received. Though it does not necessarily indicate that the argument on the short-term benefits has more ground to stand on than its counterpart, it certainly shows the positive side of TD as the practitioners collectively value the short-term gains. However, the practitioners are concerned about the negative effects of TD and its lifecycle in software development (T1.7). 

On the topic \textbf{[T2] Managing TD: A Team-Centric Approach}, practitioners have overall positive opinions on solving TD through team collaboration (T2.1), tools and strategies for reducing TD (T2.2) and understanding and managing TD as a team (T2.3). It is interesting to notice that team collaboration is one of the positive ways to solve TD and manage its effect. Adopting proper tools and strategies shall also contribute to the reduction of TD within the TD management practices. Furthermore, enhancing the collective team's understanding of TD is another perspective on which practitioners have positive opinions. On the other hand, practitioners have negative opinions regarding the teams' roles when perceiving TD (T2.4). The unclear definition of roles related to TD management could cause issues.

\textbf{[T3] Paying down TD in Agile Sprints} is also one of the main topics discussed heavily, as it is the corresponding metaphor for solving TD. Therein, allocating resources for TD in agile sprints (T3.1) as an important practice for paying back TD is positively received. So it is with the adoption of daily focus on TD in agile teams (T3.4) and TD prioritization in sprint backlog (T3.5). These "keys to success" for paying down TD in agile sprints are commonly recommended by practitioners and likely discussed positively. However, addressing TD in sprint backlogs (T3.3) is usually discussed as an issue that most likely results in negative opinions. Regarding prioritizing TD in sprint planning (T3.2), the opinions are disputed. The practitioners will likely present the steps and actionable points in this matter more neutrally.

Regarding \textbf{[T4] Understanding the Business Impact of TD}, the two negative perspectives therein are the risk assessment of TD in terms of business (T4.1) and the impact of TD on team morale and productivity (T4.3). It is reasonable to consider that when talking about potential risks caused by TD, the tone of practitioners is mostly negative. Since TD can negatively influence team morale and productivity \cite{besker2020influence}, the opinion on this subtopic is also reasonably negative. Surprisingly, practitioners' opinions are mostly positive when talking about the cost of TD on business development (T4.2). The reason is likely that practitioners collectively mention the cost of TD is a positive sign in terms of business impact. Meanwhile, evaluating the business risks of TD (T4.4) is neutral and disputed, as approximately half of the content is on listing risks and the other half on proposing how to mitigate them.

Regarding \textbf{[T5] Measuring TD: Tools and Metrics}, tracking TD with code editors and tools (T5.1) and using metrics to measure and track TD (T5.3) are both positively received by practitioners. These two topics include adopting tools and metrics to evaluate and measure TD, which can help solve its negative effects. Therefore, it is reasonable that such topics are positive. On the other hand, measuring TD with SonarQube (T5.2) is perceived negatively by practitioners, which reflects that such a function in SonarQube can be problematic. Furthermore, two other sub-topics on analyzing TD levels and violations (T5.4) and refactoring and managing TD with SonarQube (T5.5) are also negative. It is worth noticing that both features of TD measuring and refactoring with SonarQube are perceived negatively. Meanwhile, it is also reasonable that the discussion on TD violation has mostly negative tones. 

\textbf{[T6] TD in Product Development: A Necessary Evil?} has four different sub-topics. Therein, practitioners are somewhat pessimistic regarding the balance between quality and speed in product development (T6.1), as this sub-topic is perceived negatively. Meanwhile, the general notion of TD from a development perspective (T6.3) is also negative. However, considering the trade-off of TD in fast-paced markets (T6.2), practitioners are positive, which shows that the benefit of introducing TD for the short-term gain targeting the fast-paced market is well recognized as a good practice. Furthermore, the general notion of managing TD in product development teams is also positive. 

Regarding the topic \textbf{[T7] Balancing New Features and TD in Software Development}, the only negative sub-topic is regarding handling TD in continuous software development (T7.1). It reflects that the practitioners consider such a practice insufficiently considered therein. On the other hand, the other sub-topics, e.g., balancing perfection and practicality in agile development (T7.1), managing TD with feature development (T7.2), prioritizing fixes to manage TD (T7.3), and integrating new features while managing TD (T7.4) are all received positively by the practitioners, which reflects these are the practices positively recommended. 

Finally, regarding the topic \textbf{[T8] Code Quality and TD}, both data metrics and testing in managing TD (T8.1) and the impact of testing on TD in software architecture (T8.6) are received negatively. It reflects that these practices are poorly conducted in managing TD when code quality can suffer. Meanwhile, the other sub-topics on testing practices on TD, e.g., testing as a tool for quality assurance in software development (T8.3), testing and refactoring in microservice architectures (T8.4), improving code quality through testing and review (T8.5), are received positively. It reflects that testing, reviewing, and refactoring are the recommended practices to enhance code quality. Automated testing in reducing TD (T8.2) received disputed opinions. 

\section{Discussion}
\label{sec:Discussion}
Our work adopted an NLP-oriented approach to investigate the practitioners' collective opinions on TD. To this end, we collected articles and posts from three majorly popular technical forums and social media platforms (i.e., Medium, Stack Overflow, and DZone) for technical discussions. Using GSDMM topic modeling on properly preprocessed textual data, we extracted eight main topics about TD that practitioners are discussing.  


\begin{keyTakeAways}[Common aspects of TD (RQ$_1$)] \textit{The true cost of TD, team-centric management approaches, paying down TD in agile sprints, understanding its business impact, tools and metrics for measuring TD, its role in product development, dealing with TD in agile development, and the role of testing in code quality.}
\end{keyTakeAways}

\noindent Moreover, by further modeling the topics (using GSDMM) within the set of textual data associated with each main topic, we obtained the set of sub-topics for each, and by using the VADER sentiment analysis method, we can summarize if any of these sub-topics are positive or negatively perceived by a collective of practitioners.

Table \ref{tab:posnegsub} presents the potential benefits (i.e., the positive sub-topics) in terms of each main topic and the potential issues (i.e., the negative sub-topics) for each topic marked respectively in green and red color. Such results answer respectively RQ$_2$ and RQ$_3$.

In particular, on RQ$_2$ (negative sub-topics) and referring to Table  \ref{tab:posnegsub}, several practitioners reported feeling frustrated by some recurring issues in managing TD. Of those, the most important are unclear roles and a lack of involvement of development teams in TD management. In this respect, when the roles are not well-defined or, more generally, developers are not actively involved in any decision related to TD. In most cases, this gives birth to misunderstandings and ineffective TD management strategies.

Another critical complaint is the overemphasizing of business risk at the expense of technical. Suppose the management of TD mainly pays attention to reducing business risks and does not show much interest in technical debts. In that case, this may demotivate the development teams for the long-term health of the code.

Furthermore, new features have always been pushed without adequate regard for the issues about existing TD compounds built up over time. This practice slows future development cycles and increases maintenance efforts since a code base will grow in complexity and fragility if more TD is added.

Inadequate tools to track and measure TD further highlight this issue. According to developers' opinion, teams can't keep the code quality and prioritize the debt's repayment if they just rely on SonarQube to track TD metrics. Finally, balancing quality with velocity while paying less attention to integrating agile practices in general results in too much TD accumulation and deterioration of code quality.


\begin{keyTakeAways}[Common complains on TD (RQ$_2$)] \textit{Developers complain about recurring issues with TD management and role definition. Overemphasis on business risk and unbalance between pushing new features quality and quality. Inadequate tools and pressure on handling TD under agile development.}
\end{keyTakeAways}

Finally, concerning RQ$_3$, the benefits associated with active managing and paying down TD are more significant for software development teams—quick delivery and fixes to the code base. Good TD management encourages collaboration at a high level and common ground among the team members. Therefore, all stakeholders, such as engineers, managers, and other key roles, should be included in maintaining sustainable development practices that align with the goal and quality objectives. Therefore, managing TD in the product development team, instead of the commercial team or boards, allows for a more concrete view of the TD status.

Moreover, handling TD regularly in Agile can keep development teams interested and motivated. No issues are then stashed away, and the development runs at an even, higher-producing speed with increased satisfaction within teams.

Finally, a more sustainable and productive software development life cycle can be realized by balancing the need for speed and the ability to keep high code quality with structured, tool-supported methods that allow fast and effective code testing, review, and QAS processes.

\begin{keyTakeAways}[Benefits of paying TD (RQ$_3$)] \textit{Developers think that active TD management at the product development team helps to ground the issues. Furthermore, tools and comprehensive TD pay-off help automatically balance new features and code quality while delivering in an agile scenario.}
\end{keyTakeAways}
\section{Threats to Validity}
\label{sec:TV}
In this section, we discuss these threats and the strategies we adopted to mitigate them, based on the standard checklist for validity threats proposed in~\cite{WohlinExperimentation}.

\vspace{2mm}
\textbf{Construct validity}.
Construct validities are concerned with issues that to what extent the object of study truly represents the theory behind the study~\cite{WohlinExperimentation}. The RQs and the classification schema adopted might suffer from this threat. To limit this threat, the authors reviewed independently and then discussed collaboratively RQs and the related classification schema.

\vspace{2mm}
\textbf{Internal Validity}. The source selection approach adopted in this work is described in Section~\ref{sec:EmpiricalStudy}. To enable the replicability of our work, we carefully identified and reported bibliographic sources adopted to identify the peer-review literature, search engines, adopted for the gray literature,  search strings as well as inclusion and exclusion criteria.

Possible issues in the selection process are related to the selection of search terms that could have led to a non-complete set of results. 
To overcome to mitigate this risk, we applied a broad search string. This was possible because of the novelty of the topic.
To overcome the limitation of the search engines, we queried the academic literature from eight bibliographic sources, while we included the gray literature from Google, Medium Search, Twitter Search, and Reddit Search. Additionally, we applied a snowballing process to include all the possible sources.

The application of inclusion and exclusion can be affected by researchers’ opinions  and experiences. To mitigate this threat, all the sources were evaluated by at least two authors independently. 

\vspace{2mm}
\textbf{External Validity.}
External validity is related to the generalizability of the  results of our multivocal literature review.  In our study we map the literature on Open Tracing Tools, considering both the academic and the gray literature. However, we cannot claim to have screened all the possible literature, since some documents might have not been properly indexed, or possibly copyrighted or, even not freely available.

\vspace{2mm}
\textbf{Conclusion validity}.
Conclusion validity is related to the reliability of the conclusions drawn from the results~\cite{WohlinExperimentation}.
To ensure the reliability of our treatments, the terminology adopted in the schema has been reviewed by the authors to avoid ambiguities. All primary sources were reviewed by at least two authors to mitigate bias in data extraction and each disagreement was resolved by consensus, involving the third author.

\section{Related Work}
\label{sec:RW}
TD is a metaphor that reflects technical compromises that ultimately negatively affect source code maintenance; various previous studies have investigated its impact \cite{ernst2015measure,lenarduzzi2019technical} and its evolution and management \cite{li2015systematic}. For instance, \citet{li2015systematic} surveys 94 studies to systematically map the debt type, highlighting ambiguity in the broad definition of TD. Therefore, \citet{ernst2015measure} investigated whether the technical debt metaphor goes beyond the ambiguity found by \citet{li2015systematic}. Earlier studies had dealt mainly with code metrics and small developer surveys. Building on this, \citet{ernst2015measure} surveyed 1,831 participants: software engineers and architects dealing with long-term and software-intensive projects from three large organizations. They conducted follow-up interviews with seven software engineers. While the metaphor is practical for communication, the existing tools cannot handle all its details. Based on their findings, the authors propose a technical debt timeline to improve management and tooling strategies.

Similarly, recent studies survey the opinion and technical experience of developers on the TD. For instance, 
\citet{rocha2017understanding},  present the results of an online survey with 74 participants from the Brazilian software industry, aiming to understand why technical debt is introduced, eliminated, and managed, mainly at the code level. In particular, their survey goes beyond explicit technical debt management and includes poor code introduced unintentionally, which is then recognized as debt while the software evolves. The main reasons for technical debt are overload in the workload, time pressure, and pressure from management. However, participants also relate technical debt to inexperience when judging other developers. 

\citet{ramavc2022prevalence} investigated TD by preparing and conducting a family of surveys. Several questionnaires were launched within the framework of the InsighTD initiative by a 12-country research team. The questionnaires were replicated in entire national teams with industry practitioners. The authors received 653 valid responses from six countries. The results show that 22\%  of practitioners have theoretical knowledge about TD, while 47\% have practical experience. More senior practitioners working in organizations or teams with more people are much more aware of TD management. The most important root cause of TD is time pressure, whereas among its effects are time delays in delivery, low maintainability, and rework.

\citet{spinola2013investigating}  extracted a list of statements about technical debt from various online forums, blogs, and literature. Among these, 14 statements were chosen and assessed through two surveys, to which 37 practitioners answered the questionnaires. With both consensus and agreement ranking, it became evident that TD is far from just labeling "bad code" in software project management—it is essential.

Our work differentiates from previous related work on surveying developer opinions \cite{ernst2015measure,rocha2017understanding,ramavc2022prevalence,spinola2013investigating} for the scope, the methodology, and size of the surveyed opinions. More specifically, concerning \citet{spinola2013investigating} we analyzed 2,213 posts and online articles in the grey literature. Moreover, our study employed two novel methodological choices for the related work: the use of LLMs for data synthesis and the broader scope. We aimed to investigate, collect, and classify developers' opinions into major topics to provide a comprehensive and concise view of possible new research opportunities and practitioners' suggestions for consolidating good practices while averting non-optimal ones.
\section{Conclusion}
\label{sec:Conclusion}
Our study has employed an LLM-focused methodology to investigate the multivocal topic of TD in the grey literature survey as perceived by practitioners in software development. We identified eight major topics about TD from the discussion data, which included its actual cost, approaches to managing TD, how it is combined with agile practices, its business impacts, measurement tools, and its place in product development and code quality matters. 

Among the top and continuous TD management challenges that practitioners have been grumbling about during all these years are unclear team roles and insufficient developer engagement, thereby affecting proper decision-making and strategy implementation. 

On the other hand, an active management and pay-down of TD is not very cheap. It facilitates collaboration among its members and major stakeholders toward a collective approach to sustainable development practices. 

Future research efforts should focus on AI-driven TD detection and management, as well as an analysis of the effect of TD on current software development methodologies such as DevOps.

\bibliographystyle{elsarticle-num-names}
\bibliography{reference.bib}


\end{document}